\documentclass[final,11pt,authoryear]{elsarticle}
\usepackage[margin=2cm]{geometry}% by courtesy of Mico

\usepackage{graphicx}

\usepackage{textcomp}

\usepackage{amsfonts,amssymb,amsmath,amsthm,bm,cancel,mathtools,mathrsfs,upgreek}

\usepackage{makecell}

\usepackage{array}
\usepackage{multirow}

\usepackage{tabularx}
\usepackage{caption}
\usepackage{booktabs}
\usepackage{float}
\usepackage{placeins}

\usepackage{enumitem}
\usepackage{rotating}

\usepackage[table]{xcolor}
\definecolor{darknavy}{RGB}{0,51,102}

\usepackage[hidelinks,colorlinks=true,urlcolor=darknavy,linkcolor=darknavy,citecolor=darknavy]{hyperref}

\AtBeginDocument{%
  \hypersetup{citecolor=darknavy,urlcolor=darknavy,linkcolor=darknavy}%
}

\usepackage[mathlines]{lineno}

\usepackage{pifont}

\modulolinenumbers[5]

\makeatletter
\newcommand\branchhead{\@startsection{paragraph}{4}{\z@}%
           {12\p@ \@plus 6\p@ \@minus 3\p@}%
           {\p@}%
           {\normalfont\normalsize\itshape}}
\makeatother

\begin{document}

\begin{frontmatter}

\title{Composition-dependent nonlinear viscoelastic--viscoplastic behavior and constitutive framework for digitally mixed polymers spanning the glass transition}

\author[a]{Beijun Shen}
\author[a]{Mary C. Boyce\corref{cor1}}
\ead{mb3814@columbia.edu}

\cortext[cor1]{Corresponding author}

\affiliation[a]{organization={Department of Mechanical Engineering},
              addressline={Columbia University},
              city={New York},
              state={NY},
              postcode={10027},
              country={USA}}

\begin{abstract}

Multi-material PolyJet printing produces voxel-scale digital mixtures of an elastomeric photopolymer (Agilus) and a glassy photopolymer (Vero), giving a material family whose room-temperature response ranges from elastomeric to glassy. The response depends strongly on composition and loading rate, and a unified description spanning the entire family has remained challenging. Here, large-deformation uniaxial compression over nearly three orders of magnitude in strain rate reveals a nonlinear, rate-dependent load--unload response that evolves continuously with composition, from recoverable elastomeric hysteresis to glassy yield with post-yield softening, hardening, and substantial residual strain. Dynamic mechanical analysis (DMA) shows that each mixture has a single glass transition temperature ($T_g$) that shifts to higher temperature with both Vero fraction and frequency, so composition acts much as temperature or loading rate does. This time--composition equivalence motivates one constitutive structure for the family rather than a separate property set per mixture. The structure combines an equilibrium hyperelastic network with three non-equilibrium, rate-dependent branches carrying reptational, intermolecular, and glassy resistance, as the material transitions from above $T_g$ to through $T_g$ to below $T_g$. Its properties are anchored at the two endpoints (i.e., above and below $T_g$) and interpolated by smooth, physically-motivated composition scaling laws. The model captures the measured compression response of all seven calibrated compositions and predicts a withheld mixture from the scaling laws alone, as well as the plane-strain compression and uniaxial tension responses of three representative compositions. Resolving the predicted stress into its branches shows the load passing from the elastomeric to the glassy mechanism as Vero content and rate rise. The contribution to the work done captures the transition in elastic storage vs dissipation along the same path. The glass transition that DMA maps at small strain therefore governs the large-strain deformation mechanisms. The framework provides a compact, physically-based description of digitally-mixed polymers and a predictive basis for the mechanics-guided design of functionally graded, architected multi-material structures.

\end{abstract}


\end{frontmatter}

\section{Introduction}

Multi-material photopolymer jetting (PolyJet) deposits microscopic droplets of liquid photopolymer voxel by voxel, with a feature resolution possible down to about \(14~\mu\mathrm{m}\), curing each deposited layer via ultraviolet light. Co-jetting an elastomeric base (Agilus) and a glassy base (Vero) in prescribed ratios produces a family of digitally mixed materials (DMs) whose room-temperature response ranges from a soft viscoelastic elastomer to a stiff viscoplastic glassy polymer. Voxels of any composition can therefore be placed arbitrarily within a single part. Throughout this paper, ``elastomeric'' denotes a compliant, rate-dependent response that exhibits hysteresis yet remains largely recoverable, and ``glassy'' denotes an elastically stiff response that yields and retains substantial residual strain upon unloading after yield.
This voxel-scale control of composition and spatial arrangement opens a broad composition--structure--property design space, which can be exploited in architected and auxetic structures \citep[e.g.,][]{Ryvkin2020Fault-tolerantMaterial, Hunter2022ControllingStructures, Mora20223DJetting, Su2022ABehavior, Magrini2024ControlNetworks, Fox2025ControllableComposites}, in morphing and stimuli-responsive devices \citep[e.g.,][]{Ge2014ActivePrinting, Mao2015SequentialPolymers, Guttag2015LocallyComposites, Yuan20203DBehavior, Lumpe2021ComputationalMorphing, Yuan2021VoxelProperties}, and in biomimetic and energy-absorbing composites \citep[e.g.,][]{Wang2011Co-continuousDissipation, Li2013WrinklingComposites, Rivera2020TougheningBeetle, Fox2024ExtractingMaterials, Kaynia2024Soft-LayeredStorage, Abu-Qbeitah2025ExperimentalComposites}. Realizing this potential in design and high-fidelity simulation, however, requires a constitutive description of the mechanical response dependence on composition, from viscoelastic elastomeric behavior, through the rate-dependent glass transition, to a stiff glassy polymer with viscoplastic flow.

The mechanical response of an amorphous polymer is governed by its glass transition, which sets the mobility of the molecular network. Well above the glass-transition temperature \(T_g\), molecular segments are mobile: the material is elastomeric, and large, nearly reversible deformation is governed by the entropic elasticity of the entangled network~\citep{Flory1953Principles}, with rate-dependent dissipation from mechanisms such as the reptational sliding of chains \citep[e.g.,][]{deGennes1971Reptation, DoiEdwards1986Dynamics, Bergstrom1998CONSTITUTIVEELASTOMERS, Bergstrom2001DeformationElasticity}. Well below \(T_g\), segmental rotation is hindered by intermolecular resistance, and the material is glassy: the initial response is stiff, and the material yields when applied stress assists the thermally activated rotation of chain segments \citep{Argon1973APolymers}, enabling viscoplastic flow. Yield is followed by softening as the local molecular segments rotate and rearrange, then by strain hardening as the network orients and stretches \citep[e.g.,][]{Boyce1988LargeModel, Boyce1989ONPLASTICITY, Hasan1995APolymers}.
Between these limits the behavior is ``leathery'' and strongly rate-dependent, the network and intermolecular mechanisms acting in concert, with increasing strain rate stiffening the response much as decreasing temperature does, an equivalence formalized by time--temperature superposition \citep{Williams1955TheLiquids, Ferry1980ViscoelasticPolymers}. These regimes are described within a common finite-strain framework, in which the deformation gradient of each inelastic mechanism is decomposed multiplicatively into elastic and inelastic parts and the stress additively into an equilibrium and non-equilibrium contributions \citep[e.g.,][]{Simo1987OnAspects, Boyce1988LargeModel, Reese1998AAspects}.

Within this framework, the combined elastomeric--glassy response has been captured in two physical settings depending on the molecular structure. In the first, a single amorphous polymer is carried through its glass transition by temperature: the evolving response is represented as a single network with stiffness and flow resistances which evolve through the transition \citep{Dupaix2005FinitePETG, Dupaix2007ConstitutiveTransition, Qi2008FinitePolymers, Nguyen2008ThermoviscoelasticRelaxation, Westbrook2011AProcesses}. In the second, the elastomeric and glassy responses coexist as distinct microphases of a segmented copolymer, such as a thermoplastic polyurethane or polyurea, each microphase keeping its own glass transition \citep{Qi2005Stress-strainPolyurethanes, Cho2013ConstitutivePolyurea}. In both, the change from elastomeric to glassy is driven by temperature acting on one network or by the coexistence of two microphases in a copolymer. Digitally mixed polymers present a third setting. The two chemically compatible acrylate resins co-cure and interdiffuse at the voxel scale, forming a covalently bonded material whose micron-scale features lie well below the length scale at which separate transitions resolve~\citep{Yuan2021VoxelProperties, Lee2026MechanicalAnisotropy}. Each mixture therefore shows a single, broadened glass transition~\citep{Ho2024ThermomechanicalPercolation}, whose position depends on composition, temperature, and loading rate alike; at a fixed temperature and rate, composition is what places a mixture within that transition. The same finite-strain machinery of network elasticity and glassy flow carries over, but this composition-driven transition, distinct from both temperature and microphase separation, is a setting that current models were not developed to capture.

Digital materials have been widely used as constituents of multi-material 3D printed composites and architected structures, and established constitutive models have been adopted to describe them. Structures combining stiff and soft regions have been analyzed with small-strain linear elasticity \citep{Li2013WrinklingComposites, Shen2014SimpleMetamaterials, Dalaq2016MechanicalReinforcements, Liu2020CombinationComposites}; soft-dominant materials at large strain have been represented by hyperelastic strain-energy functions, among them the Neo--Hookean \citep{Cho2016EngineeringCrystals, Li2018AuxeticMetamaterials, Fernandes2021MechanicallySponges, Liu2023MechanicalTemperatures}, Mooney--Rivlin \citep{Jiang20173DRotation}, Yeoh \citep{Su2020ScaleComposites}, and Arruda--Boyce \citep{Chen2017LatticeControl, Jiang20203DPerformance, Tee2020PolyJetApproach} forms; and the stiffer, Vero-dominant materials, which yield and retain large residual strain, by elastic--plastic laws \citep{Liu2020CombinationComposites, Su2020ScaleComposites, Tee2020PolyJetApproach, Wei2021GradientResistance, Daneshdoost2024Structure-performanceEffects}. These forms are simple and readily available, and each is fitted to one printed composition, but all have been taken to be rate-independent.

Rate dependence, by contrast, is intrinsic to these materials and shapes the response of DMs and multi-material structures \citep{Ge2014ActivePrinting, Zhang2015BiomimeticTesting, Slesarenko2016HarnessingComposites, Tao20204DPerformance, Abu-Qbeitah2025ExperimentalComposites}; the models that address it fall into two groups, each developed for one end of the family. For Vero-rich and neat Vero materials, the response has been modeled by an entropic network that stiffens as its chains orient and stretch, acting in parallel with a glassy resistance that yields and then flows. This reproduces the stiff initial response, yield, post-yield softening, and hardening \citep{Wang2011Co-continuousDissipation}, and the approach has since been applied to a range of Vero-rich DMs \citep{Lee2023ExtremeCrystals, Lee2025StiffnessPolycrystalsb, Moon2025ExtremeMaterials}. Unloading and explicit rate effects, however, are seldom examined. For Agilus-rich DMs, quasi-linear viscoelasticity with a hyperelastic backbone and a Prony series has been fitted separately to each composition: a three-term series to the tensile loading response of TangoPlus and its digital mixes up to Shore~A70 \citep{Slesarenko2018Towards3D-printing}, and a five-term series to the tensile and compressive loading and stress relaxation of Agilus and TangoPlus \citep{Abayazid2020MaterialOrientations}. A physically based model in which free chains relax by reptation describes the same tensile loading data but fails above Shore~A70 \citep{Xiang2019AMaterials}. None of these models addresses unloading, and none extends beyond Shore~A70. Complementing both, dynamic mechanical analysis (DMA) maps how the moduli and $T_g$ shift with composition \citep{Akbari2018EnhancedHinges, Meisel2018ImpactJetting, Zorzetto2020PropertiesComposites, Yuan2021VoxelProperties}: each mixture has a single glass transition whose position moves continuously across the family. This small-strain picture has not, however, been carried into a finite-strain model to predict the large-strain response.

No single framework yet describes the entire family. Each existing model is calibrated separately to each composition. The intermediate mixtures near Shore~A85--95, whose glass transition lies near room temperature, are the least studied, especially under unloading and over a wide range of rates. The viscoelastic models used for the soft compositions are accurate where they are fit, but must be refitted for each new composition rather than following from the underlying mechanisms. Together, these limitations call for a physically-based description whose material properties scale smoothly with composition.

Here, we characterize the composition-dependent and rate-dependent large-deformation behavior of this family and develop a single finite-strain constitutive framework that serves every composition, from the elastomeric to the glassy endpoint. Sections~\ref{sec:ExperimentMethods} and \ref{sec:ExperimentResults} characterize the family by small-strain DMA and by large-strain compression (loading and unloading) over nearly three orders of magnitude in strain rate. The DMA shows the glass transition shifting to higher temperature with both composition and rate, so composition plays a role analogous to that of temperature and loading rate. This \textit{time--composition equivalence} motivates one constitutive structure for the whole family, with material properties that scale with composition. Section~\ref{sec:constitutive_model} develops the framework: an equilibrium network in parallel with reptational, intermolecular, and glassy non-equilibrium mechanisms. Its properties are anchored at the Agilus and Vero endpoints and interpolated by smooth, physically motivated composition scaling laws. Section~\ref{sec:Model_Exp} evaluates the model against the compression response across all compositions and rates. It then resolves the predicted stress into its mechanisms, together with the work each stores elastically and dissipates, showing that the glass transition mapped by DMA also governs the large-strain response. Section~\ref{sec:validation} tests the composition scaling by predicting a mixture withheld from the calibration. Section~\ref{sec:states} tests the model in two states of deformation not used in the calibration, plane-strain compression and uniaxial tension. Section~\ref{sec:conclusion} summarizes the findings and outlines directions for future work.
Although the model is calibrated for the Agilus--Vero system, its concept and methodology, treating composition as an analog of temperature and scaling the material properties of a single constitutive structure between the endpoints, may guide the description of similar material families more generally.

\section{Experimental methods}
\label{sec:ExperimentMethods}

\begin{figure}[htbp]
    \centering
    \includegraphics[width=0.85\textwidth]{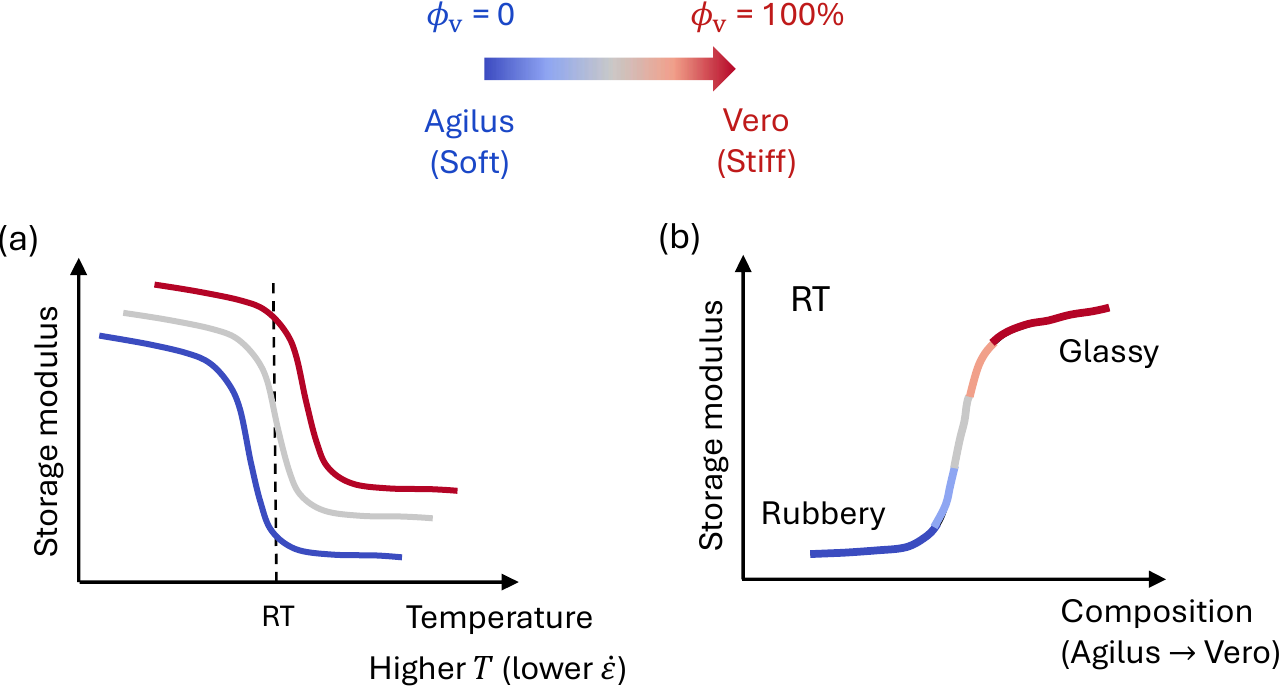}
    \caption{Schematic illustration of composition-dependent mechanical behavior in PolyJet multi-material polymers. (a) Temperature-dependent storage modulus for Agilus-rich, intermediate, and Vero-rich compositions, showing the glass transition shifting to higher temperature as the Vero fraction increases; the vertical dashed line marks room temperature (RT), where the Agilus-rich compositions are elastomeric, the Vero-rich compositions glassy, and the intermediate mixtures within the transition. (b) Reading (a) along this room-temperature line and sweeping composition gives the storage modulus versus Vero volume fraction \(\phi_V\), a continuous, S-shaped rise from elastomeric (Agilus, \(\phi_V=0\), soft) to glassy (Vero, \(\phi_V=1\), stiff).}
    \label{fig:schematic_DMA}
\end{figure}

The materials studied span the Agilus--Vero composition range, whose room-temperature response evolves from elastomeric to glassy. This composition dependence is illustrated in the sketches of the storage modulus shown in Figure~\ref{fig:schematic_DMA}.
At temperatures below their respective glass transitions, each composition is stiff and glassy, transitioning to elastomeric above the glass transition. Increasing the Vero volume fraction \(\phi_V\) shifts the glass transition to higher temperature (Figure~\ref{fig:schematic_DMA}a). At room temperature, marked by the vertical dashed line, the Agilus composition is just above its transition and is elastomeric, whereas the Vero composition is glassy, and the intermediate mixtures lie within the transition regime. Sweeping composition at this fixed temperature therefore traces a continuous, stretched "S"-shaped rise in modulus from the elastomeric to the glassy end (Figure~\ref{fig:schematic_DMA}b), suggesting that a change in composition acts much as a change in temperature does. This observation organizes the characterization and modeling that follow. This section describes the material fabrication (Section~\ref{sec:Materials}) and the characterization by dynamic mechanical analysis (Section~\ref{sec:DMA}), quasi-static uniaxial compression (Section~\ref{sec:CompressionTests_Uni}), plane-strain compression (Section~\ref{sec:PlaneStrainTests}) and uniaxial tension (Section~\ref{sec:TensionTests}).

\subsection{Materials and specimen fabrication}
\label{sec:Materials}

All specimens were 3D-printed on a Stratasys PolyJet J826 multi-material printer (Eden Prairie, MN, USA). The print heads jet droplets of liquid photopolymer voxel by voxel, and ultraviolet lamps mounted on the print carriage cure each deposited layer as the carriage passes over it. The two base resins were Agilus~30~Black (``A30''), a soft elastomer (Shore~A30) with a viscoelastic elastomeric response at room temperature, and Vero~Ultra~White (``Vero''), a stiff polymer with a glassy response.
Voxel-scale mixing of the two bases in prescribed ratios produced the digital materials using the High Mix mode with a nominal \(27~\mu\mathrm{m}\) layer thickness. The digital mixes A50, A70, A85, and A95 carry the manufacturer's nominal Shore~A labels, which rise with Vero content. A polypropylene-like grade (PP-like, RGDA-8530-DM) was also fabricated, which exhibits a predominantly glassy room-temperature response.

Stratasys does not disclose the mixing algorithm, the chemical makeup, or the base ratios of the digital grades, so their Vero volume fractions \(\phi_V\) are not precisely known. The approximate values used here draw on those reported by \citet{Slesarenko2018Towards3D-printing} and \citet{Zorzetto2020PropertiesComposites}, and are listed in Table~\ref{tab:table1} with the materials ordered by increasing stiffness. The two studies estimated the fractions independently, from the resin consumed in printing and from the voxel images sent to the printer, respectively, and agree closely where their grades overlap. Throughout, \(X\) denotes the print-head scan direction, \(Y\) the in-plane transverse direction, and \(Z\) the layer-stacking direction. Noting that print orientation may influence the measured properties \citep{Bass2016ExploringParts, Dalaq2016MechanicalReinforcements, Abayazid2020MaterialOrientations, Lee2026MechanicalAnisotropy}, each specimen geometry was printed in a single fixed orientation, stated below, for consistency across materials. Additionally, we tested tension in different orientations and found negligible anisotropy with responses in different directions overlapping one another within the variability of the responses (Section~\ref{sec:TensionTests}). After printing, support material was mechanically removed and the specimens were rinsed with water and air-dried before testing.

\begin{table}[htbp]
\centering
\caption{Materials studied in this work, ordered by increasing stiffness, with their approximate Vero volume fractions \(\phi_V\).}
\label{tab:table1}
\begin{tabular}{@{}llc@{}}
\toprule
\textbf{Category} & \textbf{Material} & \textbf{Vero volume fraction \(\phi_V\) (\%)} \\
\midrule
Agilus family     & A30      & 0    \\
\midrule
Digital materials & A50      & 8.3  \\
                  & A60$^{*}$ & 13.2 \\
                  & A70      & 18.1 \\
                  & A85      & 25.3 \\
                  & A95      & 36.2 \\
                  & PP-like  & 66 \\
\midrule
Vero family       & Vero     & 100  \\
\bottomrule
\multicolumn{3}{@{}l@{}}{\footnotesize $^{*}$\,Withheld from model calibration and used to validate the model's predictions (Section~\ref{sec:validation}).}
\end{tabular}
\end{table}

\subsection{Dynamic mechanical analysis}
\label{sec:DMA}

Dynamic mechanical analysis (DMA) characterized the small-strain viscoelastic response of each material across a wide temperature and frequency window covering both elastomeric and glassy regimes. Tests were performed on a DMA~850 (TA Instruments, New Castle, DE, USA) with liquid-nitrogen cooling. For each material, rectangular strips of length \(25~\mathrm{mm}\), width \(5~\mathrm{mm}\), and thickness \(1~\mathrm{mm}\) were printed lying flat in the \(X\)--\(Y\) plane, with the length along \(X\) and the thickness along \(Z\); a glossy finish was used throughout to avoid support-material cleanup on the thin \(Z\) face. The materials tested were A30, A70, A95, PP-like, and Vero.

Specimens were mounted in tensile clamps with an active gauge length of about \(10~\mathrm{mm}\) and subjected to sinusoidal strain at amplitudes \(\le 0.1\%\), to stay within the linear viscoelastic regime. Frequency sweeps at \(0.1\), \(1\), \(10\), and \(100~\mathrm{Hz}\) were combined with a temperature ramp of \(2~^\circ\mathrm{C}\,\mathrm{min}^{-1}\), preceded by a \(2~\mathrm{min}\) isothermal hold at the starting temperature. The temperature window was set per material to cover both elastomeric and glassy regimes: \(-80\) to \(100~^\circ\mathrm{C}\) for A30, \(-80\) to \(120~^\circ\mathrm{C}\) for A70 and A95, and \(-80\) to \(160~^\circ\mathrm{C}\) for PP-like and Vero. The storage modulus \(E'(T)\), loss modulus \(E''(T)\), and loss factor \(\tan\delta = E''/E'\) were recorded at each frequency. Two criteria were used to identify the glass-transition temperature \(T_g\): the peak of \(\tan\delta(T)\), and the intersection of the extrapolated glassy plateau with the tangent to the steepest descent of \(\log_{10} E'(T)\), referred to hereafter as the onset of the $E'(T)$ downturn.

Master curves were constructed separately for each material by time--temperature superposition \citep{Ferry1980ViscoelasticPolymers} at a common reference temperature \(T_{\mathrm{ref}}=23~^\circ\mathrm{C}\) (\(296~\mathrm{K}\)). Data sampled every \(1~^\circ\mathrm{C}\) were replotted as \(E'\), \(E''\), and \(\tan\delta\) against frequency on logarithmic axes, and curves at other temperatures were shifted horizontally along \(\log_{10}f\) to overlay the reference, giving a material-specific shift factor \(a_T(T)\) applied identically to all three quantities. The shifted data were lightly smoothed and expressed against the reduced frequency \(f_r = f\,a_T\).

\subsection{Quasi-static uniaxial compression}
\label{sec:CompressionTests_Uni}

Quasi-static uniaxial compression tests were performed at room temperature under displacement control on a ZwickRoell uniaxial testing machine with a \(10~\mathrm{kN}\) load cell (ZwickRoell GmbH \& Co.~KG, Ulm, Germany). Cylindrical specimens of two sizes were used: \(10~\mathrm{mm}\) diameter and height for the softer materials (A30, A50, A70), and \(6~\mathrm{mm}\) diameter and height for the stiffer materials (A85, A95, PP-like, Vero), the smaller geometry keeping reaction forces within the load-cell capacity. The surface finish was matte through A70 and glossy from A85 onward, and all cylinders were printed upright, with the cylinder axis along \(Z\), so compression was applied along the layer-stacking direction. Specimens were compressed between hardened steel platens lubricated with a thin oil layer to suppress friction-induced barreling and achieve near-homogeneous deformation, with a \(1~\mathrm{N}\) preload at the start of each test. To remove machine compliance, two reference marks on the upper and lower platens were tracked optically with a high-resolution camera (Sony A7R~V; Sony Corp., Tokyo, Japan), and the platen separation was used as the specimen displacement in all subsequent calculations. The displacement and force signals were lightly smoothed to remove optical-tracking quantization noise, with negligible effect on the stress, hysteresis, and residual strain.

Each test consisted of a single load--unload cycle under displacement control. The crosshead displacement rate was set to the target engineering strain rate multiplied by the initial specimen height, giving nominal rates of \(\dot{\varepsilon} = 0.001\), \(0.01\), \(0.1\), and \(0.5~\mathrm{s^{-1}}\).
The maximum applied engineering strain was approximately \(67\%\) (true strain \(\approx 1.1\)) for A30 through A95 and approximately \(54\%\) (true strain \(\approx 0.78\)) for PP-like and Vero, the latter limited by the load-cell capacity. A fresh specimen was used in each test, and most rates were repeated to assess reproducibility.

Engineering stress was computed from the measured force and the initial cross-sectional area, and engineering strain from the optically tracked specimen displacement and the initial specimen height. An initial Young's modulus was extracted from a linear fit over the initial \(4\%\) of true strain. True strain was defined as \(\ln\lambda\), where \(\lambda\) is the ratio of current to initial specimen height. Assuming incompressibility, the true stress equals the engineering stress multiplied by \(\lambda\).

\subsection{Plane-strain compression}
\label{sec:PlaneStrainTests}

Plane-strain compression tests were performed on three representative compositions, A30, A70 and A95, on the same testing machine, in a clamped channel die \(10.5~\mathrm{mm}\) wide that constrains one lateral direction of the specimen and leaves the other free. A30 and A70 were tested as \(10.5~\mathrm{mm}\) cubes and A95 as \(10.5\times6\times6~\mathrm{mm}\) blocks, \(6~\mathrm{mm}\) high, all loaded along the layer-stacking direction \(Z\). Mineral oil was used to lubricate all sides, and a PTFE bar between the punch and the specimen further reduced friction. Each specimen was loaded at \(0.01\) or \(0.001~\mathrm{s^{-1}}\) to a true strain of about \(0.7\) for A30, \(0.5\) for A70 and \(0.4\) for A95, and unloaded at the same rate. The specimen displacement was taken as the crosshead displacement corrected for the machine compliance, and the true strain and true stress were computed as in Section~\ref{sec:CompressionTests_Uni} assuming the incompressible relation between true and engineering stress. The mean of up to four tests at each composition and rate is reported with their range.

\subsection{Uniaxial tension}
\label{sec:TensionTests}

Uniaxial tension tests were performed on the same three compositions on an Instron testing machine with a \(50~\mathrm{N}\) load cell (Instron, Norwood, MA, USA). Dogbone specimens were printed flat, with the gauge section along \(X\) (\(0^\circ\)), at \(45^\circ\) or along \(Y\) (\(90^\circ\)). The gauge section was approximately \(5~\mathrm{mm}\) long and \(5~\mathrm{mm}\) wide, \(1~\mathrm{mm}\) thick for A30 and \(1\), \(2\) or \(3~\mathrm{mm}\) thick for A70 and A95. Each specimen was loaded and unloaded at a crosshead speed of \(0.22~\mathrm{mm\,s^{-1}}\), a nominal strain rate of \(0.01~\mathrm{s^{-1}}\) over the \(22~\mathrm{mm}\) grip separation. The axial strain was measured by video tracking of marks on the gauge section, and the true stress was computed as in Section~\ref{sec:CompressionTests_Uni}. As in Section~\ref{sec:PlaneStrainTests}, the mean of the tests at each orientation is reported with their range.

\section{The composition-driven glass transition and its mechanical signatures}
\label{sec:ExperimentResults}

This section establishes the physical basis for interpreting and modeling the large-deformation nonlinear, time-dependent behavior of these mixtures: a glass transition whose location depends on both composition and rate. Dynamic mechanical analysis (Section~\ref{sec:DMA_Results}) maps this transition at small strain; quasi-static compression (Section~\ref{sec:Compression_Results}) records the corresponding large-deformation signatures, from recoverable elastomeric hysteresis to glassy yield and flow.

\subsection{Dynamic mechanical analysis}
\label{sec:DMA_Results}

\begin{figure}[htbp]
    \centering
    \includegraphics[width=\textwidth]{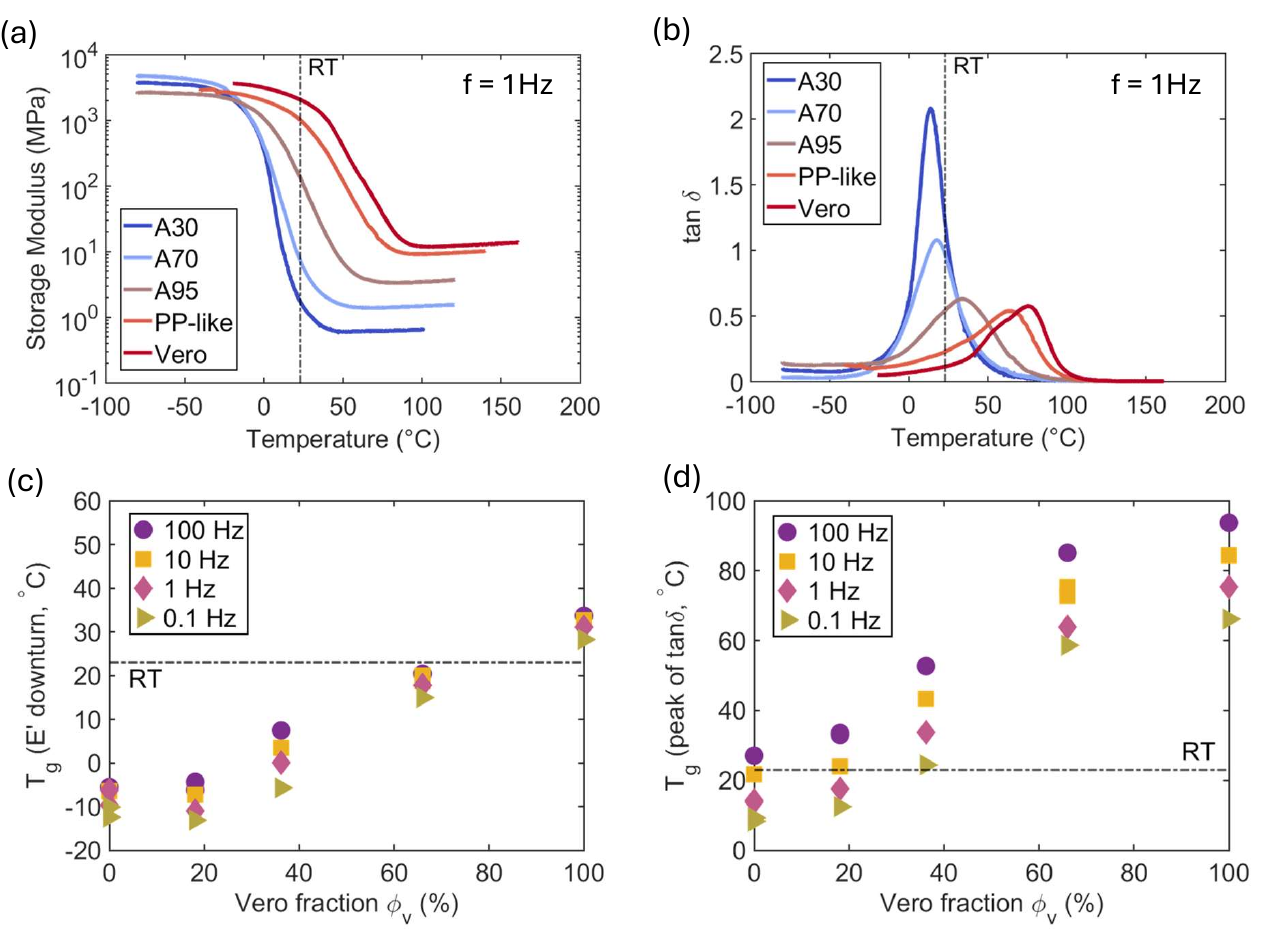}
    \caption{
    Dynamic mechanical analysis of the digitally mixed Agilus--Vero materials. (a)~Storage modulus $E'(T)$ and (b)~loss factor $\tan\delta(T)$ at $1~\mathrm{Hz}$ for representative compositions from Agilus-rich to Vero-rich; the glass-transition regime shifts systematically with composition. Glass-transition temperature $T_g$ versus the Vero volume fraction $\phi_V$ at the four tested frequencies, taken as (c) the onset of the $E'(T)$ downturn and (d) the $\tan\delta$ peak; $T_g$ rises monotonically with $\phi_V$. The room-temperature (RT) reference is $23~^\circ\mathrm{C}$.}
    \label{fig:Fig_DMA_2}
\end{figure}

Figure~\ref{fig:Fig_DMA_2}(a,b) shows the storage modulus $E'$ and loss factor $\tan\delta$ versus temperature at $1~\mathrm{Hz}$, an equivalent average strain rate of about $0.004~\mathrm{s^{-1}}$.
Below $T_g$ each material is glassy, with an $E'$ plateau of order $1~\mathrm{GPa}$ and low $\tan\delta$. On heating through the glass transition, segmental mobility increases, $E'$ falls by two to three orders of magnitude over an interval of about $40\,^\circ\mathrm{C}$, and $\tan\delta$ rises to the peak that marks the transition. Above the transition the material is elastomeric, with a low $E'$ plateau of order $10^{-1}$ to $10^{1}~\mathrm{MPa}$ and small $\tan\delta$.
As the Vero fraction increases, the glass transition shifts to higher temperature, so at room temperature ($23~^\circ\mathrm{C}$) the family spans from elastomeric A30 ($E'\approx10^{-1}$--$10^{0}~\mathrm{MPa}$) to glassy Vero ($E'\sim1$--$3~\mathrm{GPa}$). The intermediate mixtures A70 and A95 lie within the transition and are strongly frequency dependent, with PP-like nearer the glassy end.
Every composition shows a single $\tan\delta$ peak that migrates continuously with Vero content, so each mixture has one effective, composition-dependent glass transition; neat Vero alone shows a weak shoulder below its main peak.

Figure~\ref{fig:Fig_DMA_2}(c,d) reports the glass-transition temperature $T_g$ by two common measures, the onset of the $E'(T)$ downturn (c) and the $\tan\delta$ peak (d), the onset lying systematically below the peak. Both rise with Vero content and with frequency: a higher frequency shifts the transition to higher temperature, so a given material appears more glassy with increasing rate. Referenced to room temperature, A30's $T_g$ stays well below $23~^\circ\mathrm{C}$ at the lower frequencies by both measures, placing A30 in a time-dependent elastomeric state. Only at the highest frequency does its $\tan\delta$ peak rise above room temperature, as the material begins to enter its glass transition. Vero's $T_g$, by contrast, stays well above $23~^\circ\mathrm{C}$ by both measures, placing it in the glassy state. The intermediate mixtures fall on either side of room temperature, depending on composition, frequency, and which measure is used; this placement will be shown to be reflected in their large-strain behavior, with A95 showing a leathery, strongly rate-dependent compression response near the middle of the transition (Figure~\ref{fig:Zwick_compress}c). The position of each material's $T_g$ relative to room temperature is set by both composition and rate, so composition plays a role analogous to temperature and rate, a \textit{time--composition equivalence} developed further in Section~\ref{sec:gD}.

\begin{figure}[htbp]
    \centering
    \includegraphics[width=0.95\textwidth]{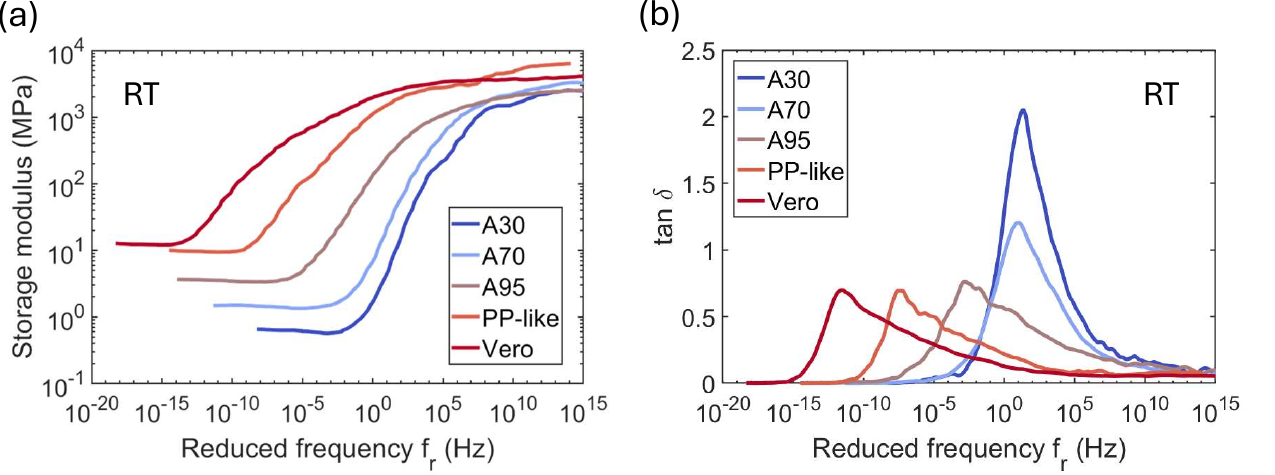}
    \caption{Master curves of (a) storage modulus $E'$ and (b) loss factor $\tan\delta$ versus reduced frequency for all tested materials, at a reference temperature of \(23~^\circ\mathrm{C}\), constructed by time--temperature superposition (Section~\ref{sec:DMA}); the transition shifts to lower reduced frequency as the Vero fraction increases.}
    \label{fig:Fig_DMA_4}
\end{figure}

Figure~\ref{fig:Fig_DMA_4} recasts the DMA data as master curves of $E'$ and $\tan\delta$ against reduced frequency at a reference temperature of $23~^\circ\mathrm{C}$, constructed by time--temperature superposition. As the Vero fraction rises, the transition moves to lower frequency and broadens, and the rubbery plateau of $E'$ climbs from $10^{-1}$--$10^{0}~\mathrm{MPa}$ for Agilus-rich materials to about $10^{1}~\mathrm{MPa}$ for Vero-rich ones, with the intermediate mixtures in between. The composition dependence of the rubbery plateau will be used to set the composition scaling of the elastomeric properties in the constitutive framework (Section~\ref{sec:calibration}). At a fixed frequency within the tested range ($0.1$--$100~\mathrm{Hz}$), each composition sits at a different point on its master curve: A30 lies on its rubbery plateau, just entering the transition, whereas higher Vero content places a material further up the transition and closer to the glassy state. The glassy modulus engaged at room temperature likewise sets the composition scaling of the glassy properties (Section~\ref{sec:calibration}).

\subsection{Quasi-static uniaxial compression}
\label{sec:Compression_Results}

\begin{figure}[htbp]
    \centering
    \includegraphics[width=\textwidth]{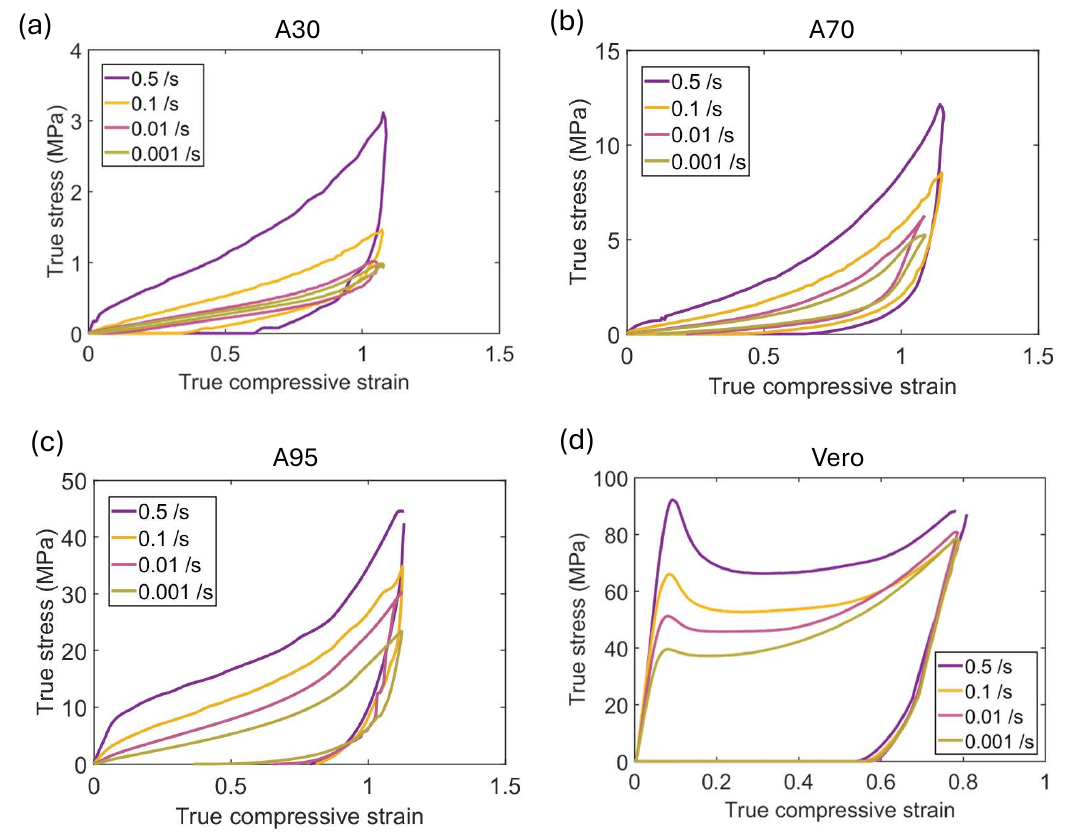}
    \caption{Uniaxial compression under load--unload cycles for four representative compositions, (a) A30, (b) A70, (c) A95, and (d) Vero, spanning the Agilus--Vero range. The response is strongly rate dependent and evolves from the largely recoverable, low-stress behavior of the Agilus-rich materials to the stiff response of Vero, which yields, dissipates throughout the load--unload cycle, and retains a large residual strain.}
    \label{fig:Zwick_compress}
\end{figure}

\begin{figure}[htbp]
    \centering
\includegraphics[width=\textwidth]{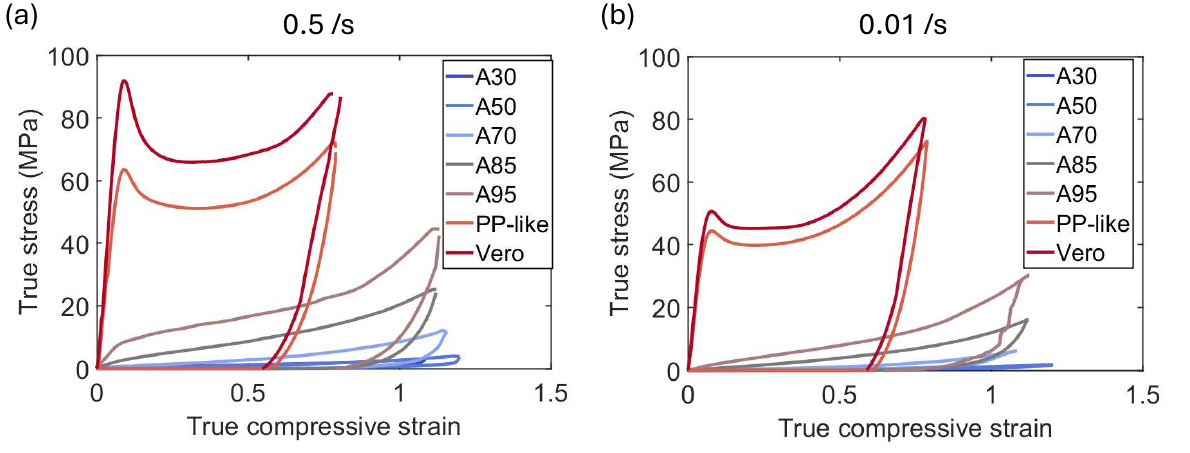}
    \caption{Overlay of true stress--true strain compression curves across the composition range at two representative rates, (a)~$\dot\varepsilon=0.5~\mathrm{s^{-1}}$ and (b)~$\dot\varepsilon=0.01~\mathrm{s^{-1}}$. Stress and stiffness rise with Vero content, and the hysteresis loops widen as the response shifts from elastomeric to glassy.}
    \label{fig:Zwick_overlay}
\end{figure}

All compositions of Table~\ref{tab:table1} were tested in quasi-static uniaxial compression at room temperature, each at engineering strain rates $\dot{\varepsilon}=0.001,\,0.01,\,0.1,$ and $0.5~\mathrm{s^{-1}}$. To limit the number of curves shown, Figure~\ref{fig:Zwick_compress} shows the true stress--true strain response of four representatives, A30, A70, A95, and Vero, that span the Agilus--Vero range, and Figure~\ref{fig:Zwick_overlay} overlays seven compositions, adding the mixtures A50, A85, and PP-like, at two representative rates. All compositions exhibit a load--unload response that is strongly nonlinear and markedly rate dependent.

The Agilus-end materials, A30 (Figure~\ref{fig:Zwick_compress}a) together with A50 and A70, are elastomeric: rate dependent yet largely recoverable.
For A30, stresses remain of order $1~\mathrm{MPa}$ even at $0.5~\mathrm{s^{-1}}$ and true strain $\approx1.1$, with nonlinear stiffening, hysteresis that increases with rate, and small residual strain that recovers fully after a short rest. The initial modulus, rollover, peak stress, and residual strain all increase with rate, whereas at the lowest rate the loading and unloading paths nearly coincide, giving a near-equilibrium response. Across A30, A50, and A70 the stress level, the hysteresis, and the immediate residual strain all increase with Vero content.
By the DMA measures these three sit above their glass transition at room temperature at the lower rates, and begin to enter the transition at the highest rate, where room temperature falls just below $T_g$ by the $\tan\delta$ peak measure.

The mixtures that lie within the room-temperature glass transition show a strongly rate-dependent rollover, from a modestly stiff, rate-dependent initial response to a more compliant elastomeric one, most pronounced at $0.5~\mathrm{s^{-1}}$ in A95 (Figure~\ref{fig:Zwick_compress}c).
This leathery response lies between the elastomeric Agilus and glassy Vero limits, each mixture's proximity to $T_g$ set by composition and rate.
At higher Vero content the materials are fully glassy: the PP-like mixture shows a stiff, rate-dependent initial slope, a clear yield, post-yield softening followed by hardening, and large hysteresis with substantial, nearly rate-insensitive residual strain (Figure~\ref{fig:Zwick_overlay}). Vero (Figure~\ref{fig:Zwick_compress}d) shows the most pronounced glassy response, with a higher, strongly rate-dependent yield stress and deeper post-yield softening before hardening.

\begin{figure}[htbp]
    \centering
    \includegraphics[width=0.95\textwidth]{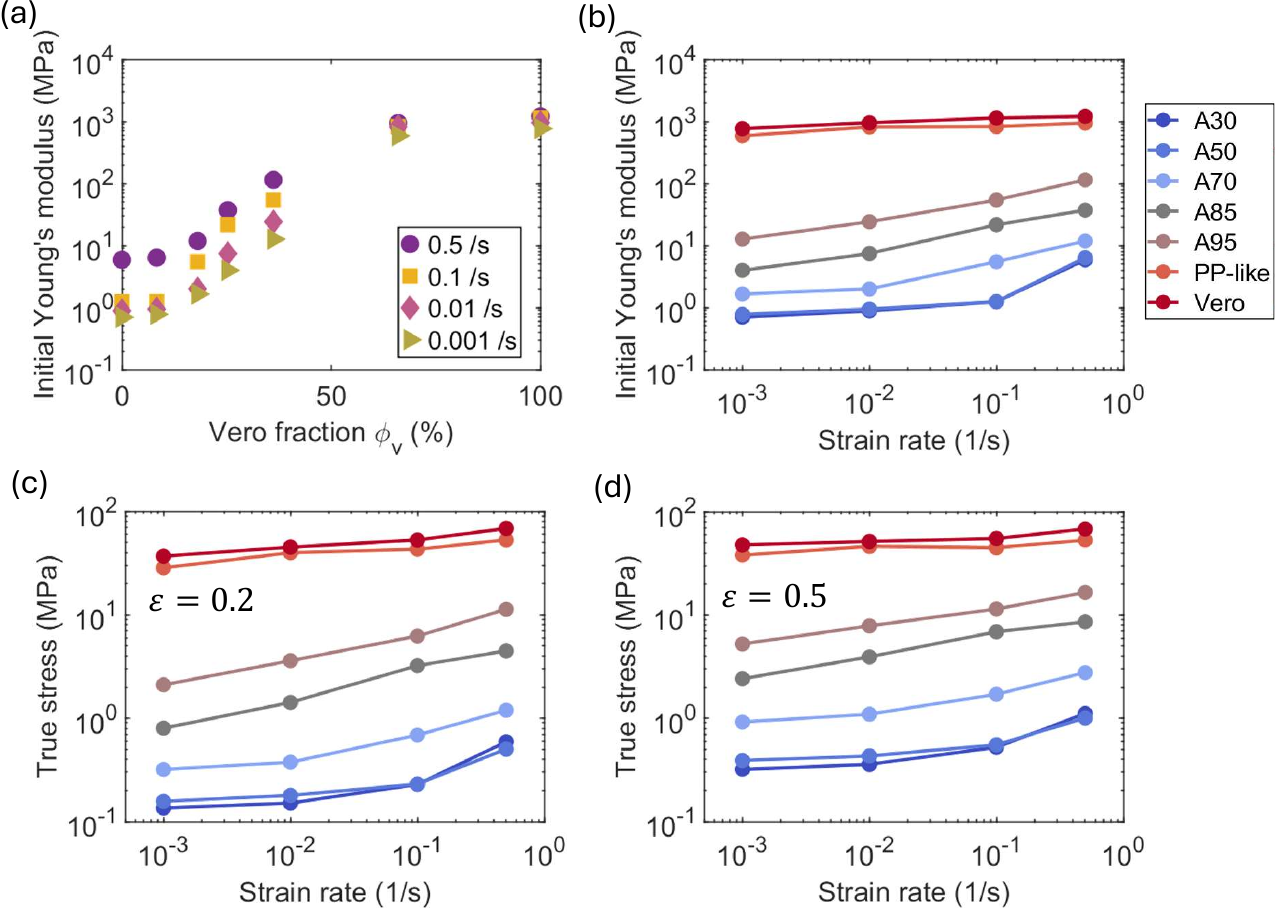}
    \caption{Rate dependence of the uniaxial compression response. (a) Initial Young's modulus, evaluated over the first $4\%$ of true strain, versus composition at the four strain rates. (b) The same moduli versus strain rate. (c,~d) True stress at true strains of $0.2$ and $0.5$ versus strain rate. In (b)--(d) the onset of strong rate sensitivity shifts to lower rates with increasing Vero content, spanning the entire rate window for A85 and A95, while PP-like and Vero remain nearly rate insensitive.}
    \label{fig:Zwick_modulus}
\end{figure}

The initial Young's modulus, evaluated over the first $4\%$ of true strain (Figure~\ref{fig:Zwick_modulus}a), rises monotonically with Vero content and is rate dependent at every composition. The modulus spans approximately three to four orders of magnitude, from $10^{-1}$--$10^{0}~\mathrm{MPa}$ for Agilus to order $10^{3}~\mathrm{MPa}$ for Vero, and its S-shaped variation with composition mirrors the room-temperature storage modulus from DMA (Figures~\ref{fig:Fig_DMA_2}a and~\ref{fig:Fig_DMA_4}a). The rate dependence of the modulus indicates an initial linear viscoelastic response, consistent with the DMA and with prior investigations of the Agilus-rich materials~\citep{Slesarenko2018Towards3D-printing, Xiang2019AMaterials, Abayazid2020MaterialOrientations}. Replotting the moduli against strain rate (Figure~\ref{fig:Zwick_modulus}b) shows the rate dependence to strengthen at a strain rate that depends on composition. For A30 and A50 the modulus is only weakly rate sensitive at the three lower rates and then rises nearly fivefold between $0.1$ and $0.5~\mathrm{s^{-1}}$, as these materials begin to enter their transition---suggesting a second, shorter-timescale relaxation mechanism. In A70 the steepening begins about a decade earlier, near $0.01$--$0.1~\mathrm{s^{-1}}$; in A85 and A95 the modulus is strongly rate sensitive over the entire tested range, consistent with their glass transitions lying at room temperature; and in PP-like and Vero the modulus is nearly rate insensitive, changing by only a factor of about $1.6$ across the window, as both remain on their glassy plateau at all tested rates. The true stress at fixed true strains of $0.2$ and $0.5$ (Figure~\ref{fig:Zwick_modulus}c,~d) shows the same transition in rate sensitivity, indicating that at least two relaxation mechanisms also govern the large-deformation behavior. The onset of the stronger rate dependence thus shifts to lower rates with increasing Vero content, from A30 through A95---the signature of two primary relaxation mechanisms for Agilus-rich materials whose governing timescales depend on composition. The compression response thus carries, at large deformation, the same composition- and rate-dependent glass transition that DMA maps at small strain. The next subsection states the time--composition equivalence that follows, and its role in motivating one constitutive structure for the whole family.

\subsection{Time--composition equivalence}
\label{sec:gD}

The DMA data organize the family through a single principle. The glass transition shifts to higher temperature with both the Vero fraction and the frequency (Figure~\ref{fig:Fig_DMA_2}c,d), so at fixed room temperature raising the Vero content moves a material through its transition just as lowering the temperature would, and raising the loading rate does the same. Composition therefore acts in an intrinsic manner on the position of the transition as temperature and loading rate do in an extrinsic manner, a time--composition equivalence analogous to the classical time--temperature equivalence. This equivalence is the physical rationale for the constitutive framework developed next. Every mixture shares one glass transition that composition displaces, so one constitutive structure describes the whole family, its properties scaled through that transition between the two endpoints.

In our recent study of the linear viscoelasticity of this family~\citep{ShenBoyceTCS}, we expressed this equivalence in closed form, as a time--temperature--composition equivalence (TTCE). Composition translates the glass transition along the frequency axis, as temperature does, but also broadens the transition and increases the rubbery plateau modulus. With a position shift, a breadth stretch and a plateau normalization, the storage moduli of the five compositions measured by DMA are found to collapse onto one master curve. The composition scaling of the constitutive model developed next uses the shift of the glass transition and the increase of the rubbery plateau modulus with Vero fraction (Figures~\ref{fig:Fig_DMA_2}c,d and~\ref{fig:Fig_DMA_4}a), both embodied in the TTCE. The material properties vary sigmoidally with composition, changing most rapidly where the shifting glass transition crosses room temperature, and the modulus of the equilibrium network follows the rubbery plateau.

\section{Finite-strain deformation constitutive framework}
\label{sec:constitutive_model}

The experimental results of Section~\ref{sec:ExperimentResults} set the requirements for the model. The DMA response locates each composition within its glass transition and gives its shift with rate (Section~\ref{sec:DMA_Results}). The compression response ranges from the time-dependent, highly recoverable elastomeric behavior of the Agilus-rich materials to the glassy, viscoplastic yield of the Vero-rich materials, with the intermediate mixtures sweeping between these limits as the rate changes (Section~\ref{sec:Compression_Results}). The model must capture both limits and properly interpolate between them in a manner consistent with the composition-dependent glass transition.

\begin{figure}[htbp]
    \centering
    \includegraphics[width=0.55\textwidth]{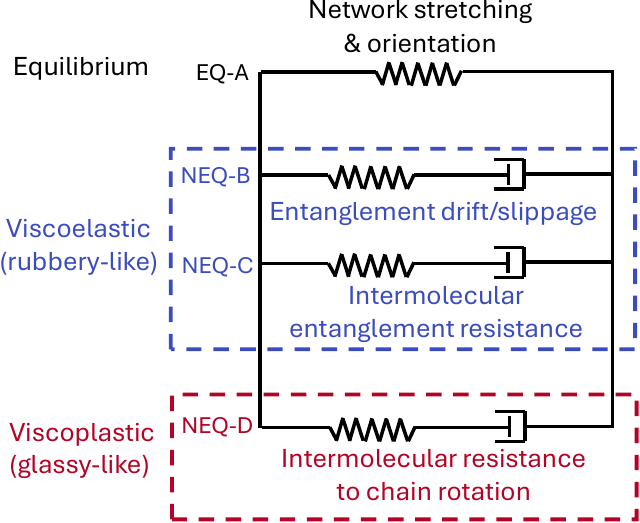}
    \caption{Schematic of the constitutive framework: an equilibrium hyperelastic network (branch~$A$) in parallel with three non-equilibrium branches, each a spring and dashpot in series. The elastomeric branches~$B$ (reptational, long-timescale) and~$C$ (intermolecular, shorter-timescale) are viscoelastic; the glassy branch~$D$ is viscoplastic. The branches contribute additively to the stress; composition enters through their moduli and flow resistances, anchored at the Agilus and Vero endpoints, and rate dependence through the flow-rule kinetics (Section~\ref{sec:constitutive_model}).}
    \label{fig:RheologicalModel}
\end{figure}

Here, a single finite-strain framework is developed that traverses every composition (Figure~\ref{fig:RheologicalModel}). An equilibrium hyperelastic network (branch~$A$) provides the rubbery backbone, acting in parallel with three non-equilibrium branches. Two non-equilibrium branches are elastomeric viscoelastic branches capturing the two primary relaxation mechanisms observed in Agilus: a long-timescale branch~$B$ that carries chain reptation within the network, and a shorter-timescale branch~$C$ that carries a developing intermolecular network resistance as the network moves up the glass transition. As an aside, we note that two primary relaxation mechanisms---one considered reptation based and one considered confinement based---have been identified in entangled polymer solutions~\citep{Zhou2018DynamicallyChains}. The third non-equilibrium mechanism is a glassy branch~$D$, evident in the uniaxial compression of Vero and well defined for glassy polymers, in which the applied stress drives chain segment rotation over the intermolecular barrier and the material flows viscoplastically. The behavior of the digital mixtures traverses these mechanisms dependent on composition and their respective glass transition. The branches contribute additively to the total Cauchy stress. Rate dependence comes from the flow-rule kinetics; composition dependence is carried by the moduli and flow resistances of the branches, which scale between the Agilus and Vero endpoints through smooth composition scaling laws (Section~\ref{sec:calibration}). This scaling rests on the physical picture established by the DMA (Section~\ref{sec:DMA_Results}): every mixture has one glass transition set by its composition (Figures~\ref{fig:Fig_DMA_2} and~\ref{fig:Fig_DMA_4}), and its properties scale accordingly.

Section~\ref{sec:kinematics} sets out the finite-strain kinematics. Section~\ref{sec:constitutive} develops the constitutive model mechanism by mechanism and assembles the total free energy and stress. Section~\ref{sec:calibration} determines the material properties, anchoring them at the Agilus and Vero endpoints and then scaling them by composition.

\subsection{Kinematics}
\label{sec:kinematics}

Let $\bm{X}$ denote a material point in the reference configuration $\mathcal{B}_0$, and $\bm{x}=\bm{\chi}(\bm{X},t)$ its position in the current configuration $\mathcal{B}_t$. The deformation gradient and velocity gradient are
\begin{equation}
\bm{F}(\bm{X},t)=\frac{\partial \bm{x}}{\partial \bm{X}},
\qquad
\bm{L}=\dot{\bm{F}}\bm{F}^{-1}=\bm{D}+\bm{W},
\label{eq:F_L_def}
\end{equation}
with rate-of-deformation $\bm{D}=\tfrac{1}{2}(\bm{L}+\bm{L}^{\top})$ and spin $\bm{W}=\tfrac{1}{2}(\bm{L}-\bm{L}^{\top})$.

To separate volume change from distortion, the deformation gradient is decomposed into volumetric and isochoric parts \citep{Simo1987OnAspects}. The isochoric (distortional) part is
\begin{equation}
\bar{\bm{F}}=J^{-1/3}\bm{F},
\qquad
J=\det\bm{F},
\qquad
\det\bar{\bm{F}}=1,
\label{eq:F_voldev}
\end{equation}
where $J$ is the volume ratio and the unimodular tensor $\bar{\bm{F}}$ carries the volume-preserving deformation. The isochoric left Cauchy--Green tensor and the effective chain stretch follow as
\begin{equation}
\bm{B}=\bm{F}\bm{F}^{\top},
\qquad
\bar{\bm{B}}=\bar{\bm{F}}\bar{\bm{F}}^{\top}=J^{-2/3}\bm{B},
\qquad
\bar{\lambda}_{\mathrm{chain}}=\bigl(\tfrac{1}{3}\,\mathrm{tr}\,\bar{\bm{B}}\bigr)^{1/2}.
\label{eq:lchain_total}
\end{equation}

All branches share the same total deformation~$\bm{F}$. Volume change is carried by a single volumetric response common to the material, so the branch kinematics are written in terms of the isochoric $\bar{\bm{F}}$. Each non-equilibrium branch $i\in\{B,C,D\}$ admits a multiplicative decomposition of $\bar{\bm{F}}$ into unimodular elastic and inelastic parts,
\begin{equation}
\bar{\bm{F}}
=\bar{\bm{F}}_A
=\bar{\bm{F}}_B^{\mathrm{e}}\bar{\bm{F}}_B^{\mathrm{v}}
=\bar{\bm{F}}_C^{\mathrm{e}}\bar{\bm{F}}_C^{\mathrm{v}}
=\bar{\bm{F}}_D^{\mathrm{e}}\bar{\bm{F}}_D^{\mathrm{v}},
\qquad
\det\bar{\bm{F}}_i^{\mathrm{e}}=\det\bar{\bm{F}}_i^{\mathrm{v}}=1,
\label{eq:F_split_branches}
\end{equation}
where $\bar{\bm{F}}_i^{\mathrm{v}}$ maps $\mathcal{B}_0$ to a stress-free intermediate configuration and $\bar{\bm{F}}_i^{\mathrm{e}}$ maps it to $\mathcal{B}_t$; the equilibrium network~$A$ deforms with $\bar{\bm{F}}$ itself.
Each branch then has its isochoric elastic left Cauchy--Green tensor and elastic chain stretch,
\begin{equation}
\bar{\bm{B}}_i^{\mathrm{e}}=\bar{\bm{F}}_i^{\mathrm{e}}\bar{\bm{F}}_i^{\mathrm{e}\top},
\qquad
\bar{\lambda}_i^{\mathrm{e}}=\bigl(\tfrac{1}{3}\,\mathrm{tr}\,\bar{\bm{B}}_i^{\mathrm{e}}\bigr)^{1/2}.
\label{eq:Bbar_branch}
\end{equation}
The isochoric velocity gradient $\bar{\bm{L}}=\dot{\bar{\bm{F}}}\bar{\bm{F}}^{-1}$, common to all branches, decomposes within each branch into elastic and inelastic parts,
\begin{equation}
\bar{\bm{L}} = \bar{\bm{L}}_i^{\mathrm{e}} + \widetilde{\bm{L}}_i^{\mathrm{v}},
\qquad
\widetilde{\bm{L}}_i^{\mathrm{v}}
= \bar{\bm{F}}_i^{\mathrm{e}}\,\dot{\bar{\bm{F}}}_i^{\mathrm{v}}\bar{\bm{F}}_i^{\mathrm{v}^{-1}}\,\bar{\bm{F}}_i^{\mathrm{e}^{-1}},
\label{eq:LiSplit_detail}
\end{equation}
where $\bar{\bm{L}}_i^{\mathrm{e}}=\dot{\bar{\bm{F}}}_i^{\mathrm{e}}\bar{\bm{F}}_i^{\mathrm{e}^{-1}}$ is the elastic velocity gradient. The tilde marks a quantity pushed forward to the current configuration by $\bar{\bm{F}}_i^{\mathrm{e}}$: here $\widetilde{\bm{L}}_i^{\mathrm{v}}$ is the push-forward of the intermediate-configuration inelastic velocity gradient $\dot{\bar{\bm{F}}}_i^{\mathrm{v}}\bar{\bm{F}}_i^{\mathrm{v}^{-1}}$, so both terms act in the current configuration. Since $\bar{\bm{F}}_i^{\mathrm{v}}$ is unimodular, $\widetilde{\bm{L}}_i^{\mathrm{v}}$ is trace-free. The intermediate configuration is defined only up to a rigid rotation, so we set the inelastic spin to zero, $\widetilde{\bm{W}}_i^{\mathrm{v}}=\bm{0}$ \citep{Boyce1989ONPLASTICITY}, and the inelastic velocity gradient reduces to a rate-of-deformation tensor, $\widetilde{\bm{L}}_i^{\mathrm{v}}=\widetilde{\bm{D}}_i^{\mathrm{v}}$.

\subsection{Constitutive model}
\label{sec:constitutive}

The model superposes a volumetric response and four branches acting in parallel (Figure~\ref{fig:RheologicalModel}): the equilibrium network~$A$ and the three non-equilibrium branches~$B$, $C$, and~$D$. Each branch stores energy in an isochoric elastic spring and contributes a deviatoric Cauchy stress~$\bm{\sigma}_i$, while the volume change is carried by a single volumetric response across all branches. Each branch is specified by its spring and its flow rule.

The three non-equilibrium branches share a common flow structure. With the inelastic spin set to zero (Section~\ref{sec:kinematics}), the inelastic rate of deformation of branch $i\in\{B,C,D\}$ is codirectional with its deviatoric stress $\bm{\sigma}_i$,
\begin{equation}
\widetilde{\bm{D}}_i^{\mathrm{v}}
=\dot{\gamma}_i^{\mathrm{v}}\,\bm{N}_i,
\qquad
\bm{N}_i = \frac{\bm{\sigma}_i}{\sqrt{2}\,\tau_i},
\qquad
\tau_i = \sqrt{\tfrac{1}{2}\,\bm{\sigma}_i:\bm{\sigma}_i}\,,
\label{eq:Dvi_BC}
\end{equation}
where $\tau_i$ is the equivalent shear stress and $\bm{N}_i$ the flow direction. The scalar flow rate $\dot{\gamma}_i^{\mathrm{v}}\ge0$ and the elastic spring of each branch are constitutively prescribed next.

\subsubsection{Branch A: equilibrium hyperelastic network}
\label{sec:branchA}

Branch~$A$ provides the rate-independent hyperelastic equilibrium backbone of the framework. It represents the entropic elasticity of the cross-linked and entangled network: the chains stretch and align as the network deforms, and their resistance rises steeply as chains approach finite extensibility. This response is captured by the Arruda--Boyce eight-chain network \citep{Arruda1993AMATERIALS}, whose isochoric energy is
\begin{equation}
\begin{gathered}
\bar{\Psi}_A = \mu_A N_A
\left[
\frac{\bar{\lambda}_{\mathrm{chain}}}{\sqrt{N_A}}\,\bar{\beta}_{\mathrm{chain}}
+\ln\!\left(\frac{\bar{\beta}_{\mathrm{chain}}}{\sinh \bar{\beta}_{\mathrm{chain}}}\right)
-\frac{1}{\sqrt{N_A}}\,\bar{\beta}_{A}^{0}
-\ln\!\left(\frac{\bar{\beta}_{A}^{0}}{\sinh \bar{\beta}_{A}^{0}}\right)
\right],
\\[4pt]
\bar{\beta}_{\mathrm{chain}}=\mathcal{L}^{-1}\!\left(\frac{\bar{\lambda}_{\mathrm{chain}}}{\sqrt{N_A}}\right),
\qquad
\bar{\beta}_{A}^{0}=\mathcal{L}^{-1}\!\left(\frac{1}{\sqrt{N_A}}\right),
\end{gathered}
\label{eq:psiA_ref}
\end{equation}
in which $\mu_A$ is the network shear modulus, $N_A$ the number of statistical segments per chain, which sets the locking stretch $\sqrt{N_A}$ at which the chains approach full extension, $\bar{\lambda}_{\mathrm{chain}}$ the effective chain stretch of Eq.~\eqref{eq:lchain_total}, and $\mathcal{L}^{-1}(\cdot)$ the inverse Langevin function, evaluated with the Pad\'{e} approximant of \citet{Cohen1991AFunction}. The last two terms are the first two evaluated at $\bar{\lambda}_{\mathrm{chain}}=1$, so the energy vanishes in the undeformed state. The corresponding Cauchy stress is deviatoric,
\begin{equation}
\bm{\sigma}_A
=\frac{\mu_A}{3J}\,\frac{\sqrt{N_A}}{\bar{\lambda}_{\mathrm{chain}}}\,
\mathcal{L}^{-1}\!\left(\frac{\bar{\lambda}_{\mathrm{chain}}}{\sqrt{N_A}}\right)
\mathrm{dev}\,\bar{\bm{B}},
\qquad
\mathrm{dev}\,\bar{\bm{B}}=\bar{\bm{B}}-\tfrac{1}{3}(\mathrm{tr}\,\bar{\bm{B}})\,\bm{I}.
\label{eq:sigmaA}
\end{equation}
The same network model serves as the equilibrium backbone for every composition, with the properties $\mu_A$ and $\sqrt{N_A}$ varying with composition (Section~\ref{sec:calibration}).

\subsubsection{Viscoelasticity}
\label{sec:viscoelasticity}

Branches~$B$ and~$C$ capture the rate-dependent dissipation seen in the load--unload response of the Agilus and intermediate compositions: the rate sensitivity of the initial slopes, the rollover stresses, and the large-strain tangential slopes (Section~\ref{sec:Compression_Results}). These features span nearly three orders of magnitude in strain rate. We represent them with two non-equilibrium mechanisms of different characteristic timescales. Branch~$B$, the long-timescale reptational relaxation of the entangled network, shapes the lower-rate response; branch~$C$, the shorter-timescale intermolecular resistance, shapes the higher-rate response, as the material enters the transition regime. Both share the kinematics of Section~\ref{sec:kinematics}, the codirectional flow of Section~\ref{sec:constitutive}, and the same eight-chain spring, and differ in their flow rules.

\branchhead{Branch B: elastomeric reptational mechanism}
\label{sec:branchB}

Branch~$B$ stores energy in an isochoric Arruda--Boyce eight-chain spring, the same network form as branch~$A$ but acting on its own elastic deformation, with shear modulus $\mu_B$ and locking stretch $\sqrt{N_B}$,
\begin{equation}
\begin{gathered}
\bar{\Psi}_B = \mu_B N_B
\left[
\frac{\bar{\lambda}_B^{\mathrm{e}}}{\sqrt{N_B}}\,\bar{\beta}_B^{\mathrm{e}}
+\ln\!\left(\frac{\bar{\beta}_B^{\mathrm{e}}}{\sinh \bar{\beta}_B^{\mathrm{e}}}\right)
-\frac{1}{\sqrt{N_B}}\,\bar{\beta}_B^{0}
-\ln\!\left(\frac{\bar{\beta}_B^{0}}{\sinh \bar{\beta}_B^{0}}\right)
\right],
\\[4pt]
\bar{\beta}_B^{\mathrm{e}}=\mathcal{L}^{-1}\!\left(\frac{\bar{\lambda}_B^{\mathrm{e}}}{\sqrt{N_B}}\right),
\qquad
\bar{\beta}_B^{0}=\mathcal{L}^{-1}\!\left(\frac{1}{\sqrt{N_B}}\right),
\end{gathered}
\label{eq:psiB}
\end{equation}
with the elastic chain stretch $\bar{\lambda}_B^{\mathrm{e}}$ and isochoric tensor $\bar{\bm{B}}_B^{\mathrm{e}}$ of Eq.~\eqref{eq:Bbar_branch}, normalized as in Eq.~\eqref{eq:psiA_ref}. The corresponding deviatoric Cauchy stress is
\begin{equation}
\bm{\sigma}_B
=\frac{\mu_B}{3J}\,\frac{\sqrt{N_B}}{\bar{\lambda}_B^{\mathrm{e}}}\,
\mathcal{L}^{-1}\!\left(\frac{\bar{\lambda}_B^{\mathrm{e}}}{\sqrt{N_B}}\right)
\mathrm{dev}\,\bar{\bm{B}}_B^{\mathrm{e}}.
\label{eq:sigmaB_spring}
\end{equation}
The locking stretch is that of the equilibrium network, $N_B=N_A$. Branches~$A$ and~$B$ act on the same molecular network and lock at the same chain extension.

Branch~$B$ represents the reptational relaxation of the entangled network and follows the Bergstr\"om--Boyce flow rule \citep{Bergstrom1998CONSTITUTIVEELASTOMERS}, written first in its original explicit form,
\begin{equation}
\dot{\gamma}_B^{\mathrm{v}}
=\dot{\gamma}_{B0}^{\mathrm{v}}
\left(\frac{\lambda_B^{\mathrm{v}}-1+\xi}{\xi}\right)^{c_B}
\left(\frac{\tau_B}{\tau_{B0}^{\mathrm{ref}}}\right)^{m_B},
\qquad
\lambda_B^{\mathrm{v}}=\sqrt{\tfrac{1}{3}\,\mathrm{tr}\,\bm{C}_B^{\mathrm{v}}}\,,
\label{eq:gammadot_B_BB}
\end{equation}
in which $\tau_B$ is the equivalent shear stress of the branch (Eq.~\eqref{eq:Dvi_BC}), $\bm{C}_B^{\mathrm{v}}=\bm{F}_B^{\mathrm{v}\top}\bm{F}_B^{\mathrm{v}}$ the (unimodular) viscous right Cauchy--Green tensor and $\lambda_B^{\mathrm{v}}$ the corresponding chain stretch, $\dot{\gamma}_{B0}^{\mathrm{v}}$ a reference shear rate, $\tau_{B0}^{\mathrm{ref}}$ the flow resistance in the undeformed state, $m_B>0$ the rate-sensitivity exponent, and $\xi>0$ a small regularization constant that keeps the rate finite in the undeformed state. The reptation exponent is bounded, $-1\le c_B\le 0$: as chains reptate through their entanglements and slippage junctions, the accumulated viscous stretch slows further flow.

Absorbing the stretch dependence into the flow resistance puts Eq.~\eqref{eq:gammadot_B_BB} in the form used in the remainder of this work,
\begin{equation}
\dot{\gamma}_B^{\mathrm{v}}
=\dot{\gamma}_{B0}^{\mathrm{v}}
\left(\frac{\tau_B}{\tau_{B0}(\lambda_B^{\mathrm{v}})}\right)^{m_B},
\qquad
\tau_{B0}(\lambda_B^{\mathrm{v}})=\tau_{B0}^{\mathrm{ref}}
\left(\frac{\lambda_B^{\mathrm{v}}-1+\xi}{\xi}\right)^{q_B},
\qquad
q_B=-\frac{c_B}{m_B},
\label{eq:gammadot_B}
\end{equation}
with $\tau_{B0}=\tau_{B0}^{\mathrm{ref}}$ in the undeformed state. The reptational slowdown now shows as a hardening of the flow resistance $\tau_{B0}$, producing the long relaxation tail that shapes the low-rate response and the slow recovery on unloading. Only the combination $\dot{\gamma}_{B0}^{\mathrm{v}}/(\tau_{B0}^{\mathrm{ref}})^{m_B}$ enters Eq.~\eqref{eq:gammadot_B}; we chose to prescribe $\dot{\gamma}_{B0}^{\mathrm{v}}=1~\mathrm{s}^{-1}$ and fit $\tau_{B0}^{\mathrm{ref}}$ to the data.

\branchhead{Branch C: elastomeric intermolecular mechanism}
\label{sec:branchC}

Branch~$C$ stores energy in an isochoric Arruda--Boyce eight-chain spring with shear modulus $\mu_C$ and locking stretch $\sqrt{N_C}$,
\begin{equation}
\begin{gathered}
\bar{\Psi}_C = \mu_C N_C
\left[
\frac{\bar{\lambda}_C^{\mathrm{e}}}{\sqrt{N_C}}\,\bar{\beta}_C^{\mathrm{e}}
+\ln\!\left(\frac{\bar{\beta}_C^{\mathrm{e}}}{\sinh \bar{\beta}_C^{\mathrm{e}}}\right)
-\frac{1}{\sqrt{N_C}}\,\bar{\beta}_C^{0}
-\ln\!\left(\frac{\bar{\beta}_C^{0}}{\sinh \bar{\beta}_C^{0}}\right)
\right],
\\[4pt]
\bm{\sigma}_C
=\frac{\mu_C}{3J}\,\frac{\sqrt{N_C}}{\bar{\lambda}_C^{\mathrm{e}}}\,
\mathcal{L}^{-1}\!\left(\frac{\bar{\lambda}_C^{\mathrm{e}}}{\sqrt{N_C}}\right)
\mathrm{dev}\,\bar{\bm{B}}_C^{\mathrm{e}},
\end{gathered}
\label{eq:sigmaC_spring}
\end{equation}
with $\bar{\beta}_C^{\mathrm{e}}=\mathcal{L}^{-1}(\bar{\lambda}_C^{\mathrm{e}}/\sqrt{N_C})$ and $\bar{\beta}_C^{0}=\mathcal{L}^{-1}(1/\sqrt{N_C})$, normalized as in Eq.~\eqref{eq:psiA_ref}. As in branch~$B$, the locking stretch is that of the equilibrium network, $N_C=N_A$. All three network springs therefore share one locking stretch, and the finite extensibility of branch~$C$ controls the residual strain left after unloading. Prior to any viscous flow, the three springs act together as a single eight-chain spring of modulus $\mu_A+\mu_B+\mu_C$. Once viscous flow begins, branches~$B$ and~$C$ relax through different flow rules and on different timescales to reflect their distinct relaxation times, so the share of the stress carried by each branch depends on the rate (Section~\ref{sec:mechanism}).

Branch~$C$ represents the shorter-timescale intermolecular resistance of the entangled network, the friction that neighboring chain segments exert on one another as they rearrange. Its inelastic flow rate superposes a linear (Newtonian) term, which captures the rate-dependent initial modulus, and a power-law overstress term,
\begin{equation}
\dot{\gamma}_C^{\mathrm{v}}
=\frac{\tau_C}{\eta_C}
+\dot{\gamma}_{C0}^{\mathrm{v}}
\left(\frac{\tau_C}{\tau_{C0}(\lambda_C^{\mathrm{v}})}\right)^{m_C},
\qquad
\lambda_C^{\mathrm{v}}=\sqrt{\tfrac{1}{3}\,\mathrm{tr}\,\bm{C}_C^{\mathrm{v}}}\,,
\label{eq:gammadot_C}
\end{equation}
in which $\tau_C$ is the equivalent shear stress of the branch (Eq.~\eqref{eq:Dvi_BC}), $\eta_C$ the linear viscosity, $\tau_{C0}$ the power-law flow resistance, $m_C\ge 1$ the rate-sensitivity exponent, and $\lambda_C^{\mathrm{v}}$ the viscous chain stretch of the branch, formed from $\bm{C}_C^{\mathrm{v}}$ as in branch~$B$. Both the viscosity and the resistance evolve with the viscous deformation of the branch.

The viscosity of the linear term evolves with the state of the branch. Deformation stretches and orients the segments, so the viscosity is taken to increase with the viscous chain stretch of the branch toward a saturated level,
\begin{equation}
\dot{\eta}_C
=h_{\eta}\,\eta_C^0
\left(1-\frac{\eta_C}{\eta_C^{\mathrm{ss}}}\right)
\bigl|\dot{\lambda}_C^{\mathrm{v}}\bigr| ,
\qquad
\eta_C(0)=\eta_C^0 ,
\qquad
\eta_C^{\mathrm{ss}}>\eta_C^0 ,
\label{eq:eta_evolution}
\end{equation}
in which $\eta_C^0$ is the initial viscosity, $h_{\eta}$ the rate of growth, and $\eta_C^{\mathrm{ss}}$ the saturated value. The growth rate $h_{\eta}$ and the ratio $\eta_C^{\mathrm{ss}}/\eta_C^{0}$ are taken, for simplicity, to be the same at every composition. The growth of $\eta_C/\eta_C^0$ with the viscous chain stretch is then the same for every composition, and composition changes the viscosity only through its initial value $\eta_C^0$ (Section~\ref{sec:comp_scaling}).

The resistance of the power-law term evolves with the same state variable. The stretching and orientation that raise the viscosity also harden the resistance to flow,
\begin{equation}
\tau_{C0}(\lambda_C^{\mathrm{v}})=\tau_{C0}^{\mathrm{ref}}\,\bigl[\,1+\alpha_C\,(\lambda_C^{\mathrm{v}}-1)\,\bigr],
\label{eq:hardening_C}
\end{equation}
in which $\tau_{C0}^{\mathrm{ref}}$ is the resistance in the undeformed state and $\alpha_C\ge 0$ the hardening coefficient. Branch~$B$ hardens with its own viscous stretch in the same manner (Eq.~\eqref{eq:gammadot_B}). Only the combination $\dot{\gamma}_{C0}^{\mathrm{v}}/(\tau_{C0}^{\mathrm{ref}})^{m_C}$ enters the flow rule, so the reference rate is again prescribed, $\dot{\gamma}_{C0}^{\mathrm{v}}=1~\mathrm{s}^{-1}$.

The two viscous terms act in near succession. For $m_C>1$ the power-law overstress is negligible while the stress stays well below $\tau_{C0}$, so the linear term sets the rate dependence of the initial slope. With continued deformation, the linear contribution diminishes and the power-law contribution takes over, producing the yield-like rollover and the inelastic deformation at large strain.

\subsubsection{Branch D: glassy mechanism}
\label{sec:branchD}

Branch~$D$ captures the glassy non-equilibrium response: a stiff initial slope, rate-dependent yield, post-yield softening or a plateau, and substantial residual strain upon unloading after yield (Section~\ref{sec:Compression_Results}). The branch yields at small elastic stretch, so its elastic response is essentially linear and modeled as an isochoric Neo-Hookean spring,
\begin{equation}
\bar{\Psi}_D=\tfrac{1}{2}\,\mu_D\bigl[\,\mathrm{tr}\,\bar{\bm{B}}_D^{\mathrm{e}} - 3\,\bigr],
\qquad
\bm{\sigma}_D=\frac{\mu_D}{J}\,\mathrm{dev}\,\bar{\bm{B}}_D^{\mathrm{e}},
\label{eq:sigmaD}
\end{equation}
where $\mu_D$ is the glassy shear modulus, varying with composition (Section~\ref{sec:calibration}), and $\bar{\bm{B}}_D^{\mathrm{e}}$ follows Eq.~\eqref{eq:Bbar_branch}.

Glassy flow proceeds by thermally activated rotation of chain segments over an intermolecular barrier. The applied stress biases that barrier, lowering it in the flow direction and raising it in the reverse direction, and the difference between the forward and reverse rates gives a hyperbolic sine \citep[e.g.,][]{Boyce1988LargeModel, Hasan1995APolymers}. The flow rate superposes this activated term with a linear (Newtonian) term,
\begin{equation}
\dot{\gamma}_D^{\mathrm{v}}
=\frac{\tau_D}{\eta_D}
+\dot{\gamma}_{0}^{D}\exp\!\left(-\frac{\Delta G}{k\theta}\right)
\sinh\!\left(\frac{\Delta G}{k\theta}\,\frac{\tau_D}{s}\right),
\label{eq:gammadotD}
\end{equation}
in which $\tau_D$ is the equivalent shear stress of the branch (Eq.~\eqref{eq:Dvi_BC}), $s$ the athermal shear strength, the intermolecular resistance to segment rotation, $\Delta G$ the zero-stress activation energy, $k$ Boltzmann's constant, $\theta$ the absolute temperature, $\dot{\gamma}_{0}^{D}$ a reference shear rate, and $\eta_D$ the linear viscosity, of initial value $\eta_D^0$. The branch yields once $\tau_D$ approaches $s$, at a stress that shifts with the logarithm of the applied rate. Below yield the activated rate is negligible and the linear term carries the flow. It represents the relaxation of the glassy stiffness: in the transitional compositions such as A85 and A95, where the glass transition sits at room temperature, the initial stiffness falls strongly with decreasing rate (Section~\ref{sec:Compression_Results}); at the glassy end the relaxation timescale lies beyond the tested window. The viscosity evolves by Eq.~\eqref{eq:eta_evolution}, driven by the viscous chain stretch $\lambda_D^{\mathrm{v}}$ of the branch and with the same $h_{\eta}$ and saturation ratio, so the linear viscosity is taken to increase as the viscous deformation accumulates.

The strength evolves with the accumulated inelastic shear strain to a saturated state \citep{Boyce1988LargeModel},
\begin{equation}
\dot{s}=h_{s}\left(1-\frac{s}{s_{\mathrm{ss}}}\right)\dot{\gamma}_D^{\mathrm{v}},
\qquad
s(0)=s_0,
\label{eq:s_evolution}
\end{equation}
in which $h_{s}\ge 0$ is a softening slope, $s_0$ the initial strength, set at the scale of the measured yield stress, and $s_{\mathrm{ss}}$ a saturated strength with $0<s_{\mathrm{ss}}\le s_0$. With $s_{\mathrm{ss}}<s_0$ and $h_{s}>0$ the strength decays toward $s_{\mathrm{ss}}$ and the resistance softens, producing the post-yield stress drop; with $h_{s}=0$ the strength stays at $s_0$ and the post-yield response is a nearly flat plateau.

\subsubsection{Total free energy and stress}
\label{sec:total}

The energy stored in the four branches (Sections~\ref{sec:branchA}--\ref{sec:branchD}), together with a volumetric term, gives the Helmholtz free energy per unit reference volume,
\begin{equation}
\Psi
=\bar{\Psi}_A(\bar{\bm{B}})
+\bar{\Psi}_B(\bar{\bm{B}}_B^{\mathrm{e}})
+\bar{\Psi}_C(\bar{\bm{B}}_C^{\mathrm{e}})
+\bar{\Psi}_D(\bar{\bm{B}}_D^{\mathrm{e}})
+\Psi_{\mathrm{vol}}(J),
\label{eq:Psi_total}
\end{equation}
where $\Psi_{\mathrm{vol}}(J)=\tfrac{1}{2}\kappa(J-1)^2$ is the volumetric energy and $\kappa$ the bulk modulus, set far larger than the shear moduli of the branches so that the material is nearly incompressible.

The Cauchy stress follows from the free energy of Eq.~\eqref{eq:Psi_total}. The volumetric energy gives the hydrostatic stress. The energy of each branch gives the deviatoric stress of that branch when differentiated with respect to the isochoric deformation it carries, which is $\bar{\bm{B}}$ for branch~$A$ and $\bar{\bm{B}}_i^{\mathrm{e}}$ for branches~$B$, $C$ and~$D$. These branch stresses are given by Eqs.~\eqref{eq:sigmaA}, \eqref{eq:sigmaB_spring}, \eqref{eq:sigmaC_spring} and~\eqref{eq:sigmaD}, and the total Cauchy stress is their sum,
\begin{equation}
\bm{\sigma}=\bm{\sigma}_A+\bm{\sigma}_B+\bm{\sigma}_C+\bm{\sigma}_D + \kappa(J-1)\,\bm{I}.
\label{eq:sigma_sum}
\end{equation}
The same four branches and volumetric term are present for every composition; the response evolves across the family through the composition scaling of material properties, determined in Section~\ref{sec:calibration}.

\subsubsection{Thermodynamic consistency}
\label{sec:thermo}

With the free energy of Eq.~\eqref{eq:Psi_total} and the flow rules of Sections~\ref{sec:branchB}--\ref{sec:branchD} in place, the framework can be checked against the second law. In isothermal processes the local dissipation per unit reference volume must satisfy the Clausius--Duhem inequality~\citep{ColemanNoll1963}
\begin{equation}
\mathcal{D}=\bm{\tau}:\bm{D}-\dot{\Psi}\ge 0 ,
\label{eq:CD_inequality}
\end{equation}
where $\bm{\tau}=J\bm{\sigma}$ is the Kirchhoff stress and $\bm{D}$ the rate of deformation. Each branch stress is the hyperelastic conjugate of its isochoric energy, so the volumetric and elastic rate terms cancel and only the inelastic flow of the three non-equilibrium branches survives,
\begin{equation}
\mathcal{D}
=\bm{\tau}_B:\widetilde{\bm{D}}_B^{\mathrm{v}}
+\bm{\tau}_C:\widetilde{\bm{D}}_C^{\mathrm{v}}
+\bm{\tau}_D:\widetilde{\bm{D}}_D^{\mathrm{v}} ,
\label{eq:Dissipation_decomp}
\end{equation}
with $\bm{\tau}_i=J\bm{\sigma}_i$ driving branch~$i$; the equilibrium branch~$A$ is purely elastic. The codirectional flow of Eq.~\eqref{eq:Dvi_BC} gives $\bm{\sigma}_i:\bm{N}_i=\sqrt{2}\,\tau_i$, so each term is the product of an equivalent shear stress and a flow rate,
\begin{equation}
\mathcal{D}=\sqrt{2}\,J\sum_{i=B,C,D}\tau_i\,\dot{\gamma}_i^{\mathrm{v}} .
\label{eq:Dissipation_closed}
\end{equation}
Both are non-negative. Hence $\mathcal{D}\ge 0$ for every deformation history and the framework is thermodynamically consistent, with the standard structure of finite-strain viscoelasticity and viscoplasticity \citep{Reese1998AAspects,Boyce1988LargeModel, Bergstrom1998CONSTITUTIVEELASTOMERS}.

Integrating Eq.~\eqref{eq:CD_inequality} over the deformation history splits the work supplied into a stored and a dissipated part,
\begin{equation}
W=\int_{0}^{t}\bm{\tau}:\bm{D}\,\mathrm{d}t'=\Psi+W_d ,
\qquad
W_d=\int_{0}^{t}\mathcal{D}\,\mathrm{d}t' ,
\label{eq:work_split}
\end{equation}
with $\Psi$ the free energy of Eq.~\eqref{eq:Psi_total} and $W_d$ the accumulated dissipation, expressed per unit reference volume.

\subsection{Determination of the material properties}
\label{sec:calibration}

The seven compositions share the constitutive form of Sections~\ref{sec:kinematics} and~\ref{sec:constitutive} and differ only in their material properties. The properties are determined from uniaxial compression data, anchored at the two endpoints of the family, Agilus and Vero, and their variation across the family is described by smooth composition functions to be defined in Section~\ref{sec:comp_scaling}.

Each endpoint calibrates the properties of its active mechanisms. Agilus is elastomeric at room temperature over the tested rates, and its response is carried by the equilibrium network~$A$ and the elastomeric branches~$B$ and~$C$. Vero is glassy over the same rates, and its response is carried by the glassy branch~$D$. The elastomeric properties are anchored at the Agilus end and increase with Vero content; the glassy properties are anchored at the Vero end and decrease toward the elastomeric end.

At each endpoint the properties are determined in sequence, each from the response feature it governs. The equilibrium network is determined first; the equilibrium stress is then subtracted from the total stress, leaving the non-equilibrium stress for the viscous branches. Where a branch has both a linear viscous term and a nonlinear flow term, the linear term is read from the rate dependence of the initial slope and the nonlinear term from the yield-like rollover and its shift with the logarithm of rate.

\subsubsection{Elastomeric anchor: Agilus}
\label{sec:agilus_anchor}

\begin{figure}[htbp]
    \centering
    \includegraphics[width=\textwidth]{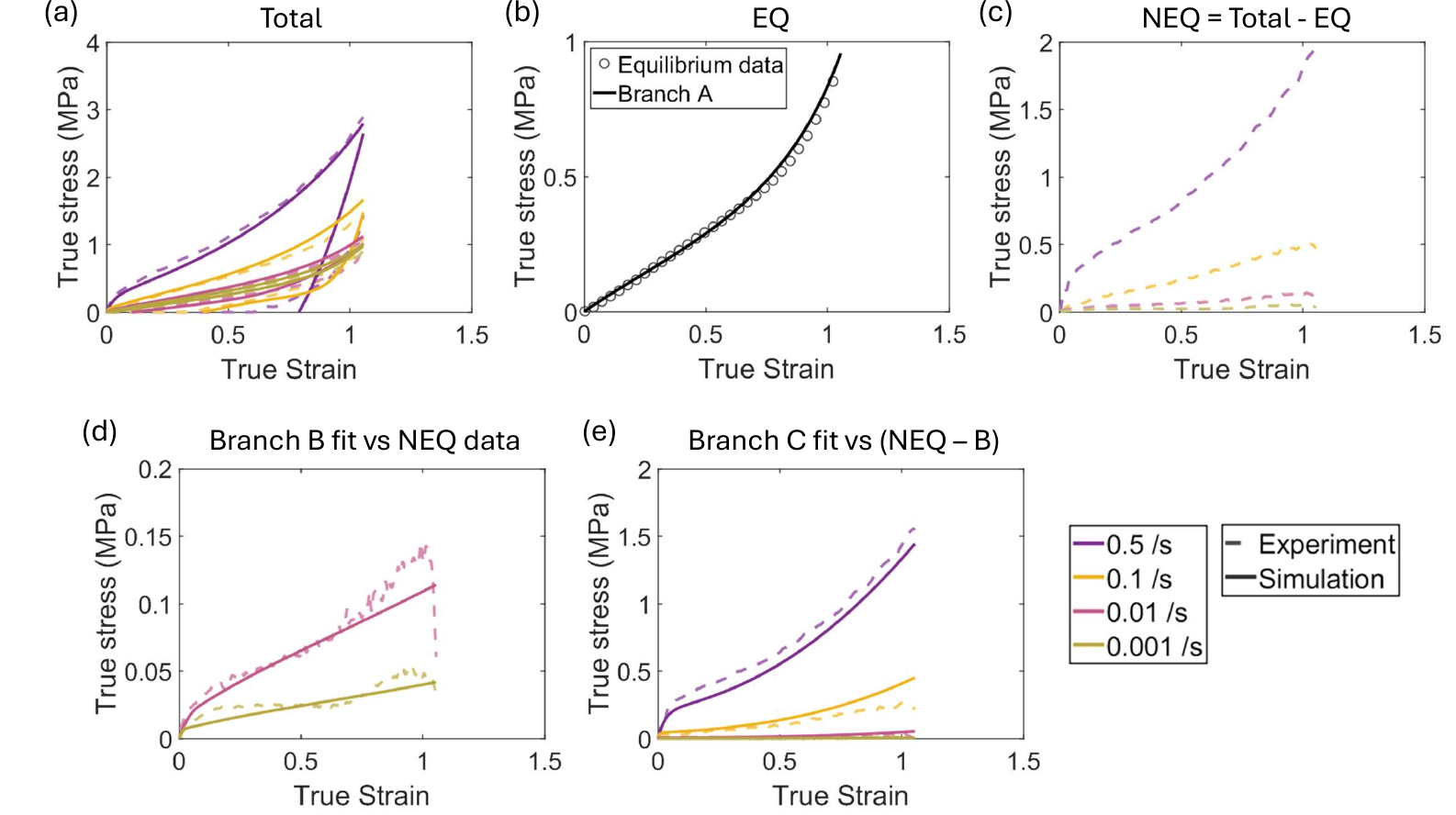}
    \caption{Determination of the elastomeric properties at the Agilus anchor. (a)~Measured and computed true stress against true strain in uniaxial compression at four strain rates (dashed, experiment; solid, model). (b)~Near-equilibrium response, obtained by averaging the loading and unloading paths at the lowest rate, with the fitted Arruda--Boyce network of branch~$A$. (c)~Non-equilibrium stress remaining once the equilibrium stress is subtracted. (d)~Branch~$B$ fitted to the non-equilibrium stress at the two lowest rates, where the shorter-timescale branch~$C$ has relaxed. (e)~Branch~$C$ fitted to the stress remaining after branch~$B$ is subtracted, at all four rates.}
    \label{fig:agilus_calib}
\end{figure}

Agilus is elastomeric at room temperature, just above its glass transition (Figure~\ref{fig:Fig_DMA_2}), and remains in the elastomeric regime across the tested rates of $10^{-3}$ to $0.5~\mathrm{s^{-1}}$. Its room-temperature master curve (Figure~\ref{fig:Fig_DMA_4}) places the glassy response near $10^{3}~\mathrm{s^{-1}}$ and above, decades beyond the tested rates. The glassy branch is therefore inactive here, and the measured response is carried by branches~$A$, $B$, and~$C$. Figure~\ref{fig:agilus_calib} shows the stages of the determination.

At the lowest rate the loading and unloading paths nearly coincide, and their average is the near-equilibrium response of branch~$A$. Its initial slope gives $\mu_A$ and its large-strain upturn gives the locking stretch $\sqrt{N_A}$ (Figure~\ref{fig:agilus_calib}b). Since $N_B=N_C=N_A$, this single fit also fixes the locking stretch of both viscous branches. Subtracting the equilibrium stress from the total stress leaves the non-equilibrium stress that the viscous branches carry (Figure~\ref{fig:agilus_calib}c).

Branch~$B$ is determined next from the data at the two lowest rates, where the shorter-timescale branch~$C$ has relaxed (Figure~\ref{fig:agilus_calib}d). The rate-dependent initial slopes give the modulus $\mu_B$. Reducing the rate-dependent rollover stress gives the exponent $m_B=2$ and the combination $\dot{\gamma}_{B0}^{\mathrm{v}}/(\tau_{B0}^{\mathrm{ref}})^{m_B}$; prescribing $\dot{\gamma}_{B0}^{\mathrm{v}}=1~\mathrm{s^{-1}}$ then gives $\tau_{B0}^{\mathrm{ref}}=0.132~\mathrm{MPa}$. The increase of the flow resistance beyond the rollover gives the reptation exponent $c_B=-0.4$, i.e., $q_B=0.2$. Branch~$C$ is then fitted to the remaining non-equilibrium stress, after subtracting branch~$B$ (Figure~\ref{fig:agilus_calib}e). The rate-dependent initial slopes provide $\mu_C$ and $\eta_C^0$; the rate-dependent rollover stress gives $\tau_{C0}^{\mathrm{ref}}$ (again with $\dot{\gamma}_{C0}^{\mathrm{v}}=1~\mathrm{s^{-1}}$) and $m_C$; and the rise of the remainder at large strain gives $\alpha_C$. The growth rate $h_{\eta}$ and saturation ratio $\eta^{\mathrm{ss}}/\eta^{0}$ of Eq.~\eqref{eq:eta_evolution} are read from the strain range over which the linear contribution fades. One pair, $h_{\eta}=80$ and $\eta^{\mathrm{ss}}/\eta^{0}=60$, is taken to serve branches~$C$ and~$D$ at every composition. The assembled model is compared with the experimental data at all four rates in Figure~\ref{fig:agilus_calib}a, and the resulting elastomeric properties are listed in Table~\ref{tab:v2_parameters}.

\subsubsection{Glassy anchor: Vero}
\label{sec:vero_anchor}

\begin{figure}[htbp]
    \centering
    \includegraphics[width=0.85\textwidth]{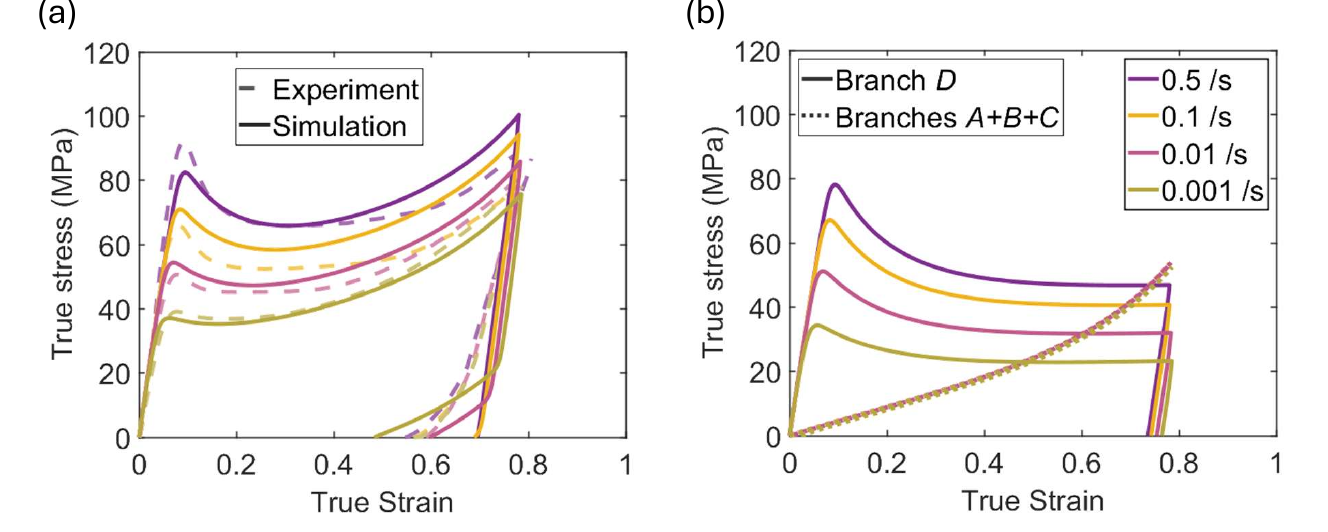}
    \caption{Determination of the glassy properties at the Vero anchor. (a)~Measured and computed true stress against true strain in uniaxial compression at four strain rates (dashed, experiment; solid, model). (b)~The computed stress resolved into the glassy branch~$D$ (solid) and the sum of the elastomeric branches $A+B+C$ (dotted). Branch~$D$ carries the rate dependence, yielding at a rate-dependent peak and softening to a rate-ordered plateau; the elastomeric sum is rate independent and hardens with the network stretch, overtaking the softened glassy stress at large strain.}
    \label{fig:vero_calib}
\end{figure}

The glassy endpoint is calibrated next, and the sequence mirrors the elastomeric one. At room temperature Vero is glassy across the tested rates (Figure~\ref{fig:Fig_DMA_2}), so the elastomeric branches~$B$ and~$C$ cannot flow on the timescale of a test and act as elastic springs. Figure~\ref{fig:vero_calib}b separates the computed response on this basis: the elastomeric sum $A+B+C$ is rate independent, while branch~$D$ carries the rate dependence and yields. Sharing the locking stretch, $N_B=N_C=N_A$, the three elastomeric springs act as one Arruda--Boyce spring of modulus $\mu_A+\mu_B+\mu_C$, whose large-strain upturn gives $\sqrt{N_A}$ and whose level gives the summed modulus. That sum is divided among the branches by the composition scaling of Section~\ref{sec:comp_scaling}, anchored at the Agilus values.

Branch~$D$ is determined from the rate-dependent part of Figure~\ref{fig:vero_calib}b, following the procedure established for glassy polymers by \citet{Boyce1988LargeModel, Hasan1995APolymers, Arruda1995EffectsPolymers}. The measurements are first converted to the scalar measures of the flow rule: $\tau=\sigma/\sqrt{3}$ for uniaxial stress $\sigma$, and $\dot{\gamma}_D^{\mathrm{v}}=\sqrt{3/2}\,|\dot{\varepsilon}|$ at the yield peak. The glassy modulus is read from the initial loading slope, giving $\mu_D=340~\mathrm{MPa}$, and the initial strength is taken at the scale of the measured yield stress, $s_0=48.3~\mathrm{MPa}$; the ratio $s_0/\mu_D=0.142$ then holds at every composition (Table~\ref{tab:v2_parameters}). The activation energy $\Delta G$ and the reference rate $\dot{\gamma}_0^{D}$ follow from the rate dependence of the yield stress, giving $\Delta G=29.5~\mathrm{kJ\,mol^{-1}}$ and $\dot{\gamma}_0^{D}=1.84~\mathrm{s^{-1}}$. A single barrier of this height serves every composition, with composition entering the flow rule through $s_0$.

The remaining glassy properties come from the shape of the response after yield: the softening parameters $h_{s}$ and $s_{\mathrm{ss}}$ from the sharpness and the depth of the post-yield stress drop through Eq.~\eqref{eq:s_evolution}, and the initial viscosity $\eta_D^0$ of the linear term from the slowest-rate curve, its evolution constants shared with branch~$C$. The assembled model captures the experimental data at all four rates (Figure~\ref{fig:vero_calib}a). We note that at the high strain rates there is heating from the plastic dissipation \citep{Arruda1995EffectsPolymers}, and the experimental curves show increased softening at large strain that the isothermal model does not represent. The glassy properties are listed in Table~\ref{tab:v2_parameters}.

\subsubsection{Composition scaling of mixtures}
\label{sec:comp_scaling}

\begin{figure}[htbp]
    \centering
    \includegraphics[width=0.9\textwidth]{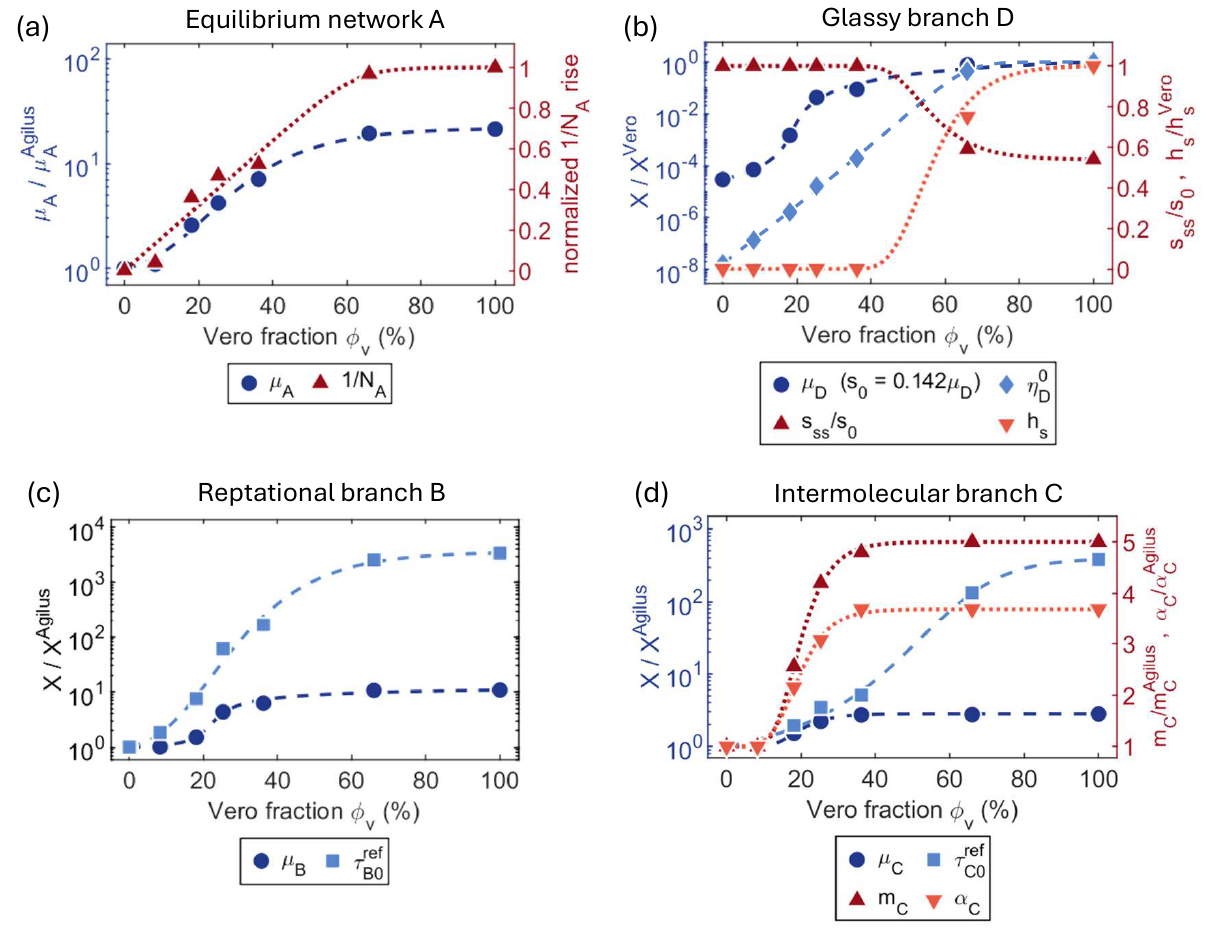}
    \caption{Composition scaling of the calibrated properties against Vero fraction $\phi_V$: (a)~equilibrium network~$A$, (b)~glassy branch~$D$, (c)~reptational branch~$B$, (d)~intermolecular branch~$C$. Markers are the per-composition values of Table~\ref{tab:v2_parameters}; curves are the fitted functions of Eq.~\eqref{eq:scaling_form} and Table~\ref{tab:scaling}, dashed for the left axis and dotted for the right. Elastomeric properties are normalized by their Agilus value, glassy properties by their Vero value. In~(b), $\mu_D$ and $s_0=0.142\,\mu_D$ share one function and fall on the same curve. In~(d), $\mu_C$ and $\tau_{C0}^{\mathrm{ref}}$ are on the logarithmic left axis and $m_C$ and $\alpha_C$ on the linear right axis, each normalized by its Agilus value, so all four start at unity; the linear viscosity $\eta_C^{0}$ is omitted, since its five-decade rise would compress the other three.}
    \label{fig:param_scaling}
\end{figure}

The calibrated properties of the Agilus--Vero mixtures are described as functions of the Vero fraction $\phi_V$. The functions share a single form built on a normalized sigmoid $\hat{S}$ of the composition, the generalized logistic of \citet{Richards1959AOf} written in $\sinh^{-1}u$,
\begin{equation}
\begin{gathered}
X(\phi_V)=X_A+\Delta_X\,\hat{S}(\phi_V),
\qquad
\hat{S}(\phi_V)=\frac{S(\phi_V)-S(0)}{S(1)-S(0)},
\\[6pt]
S(\phi_V)=\bigl[1+\nu\,e^{-k\sinh^{-1}u}\bigr]^{-1/\nu},
\qquad
u=\frac{\phi_V-\phi_V^{0}}{\delta},
\end{gathered}
\label{eq:scaling_form}
\end{equation}
applied either to the property itself or to its base-ten logarithm. Here $\phi_V^{0}$ locates the transition, $k$ sets its steepness, $\nu$ its asymmetry, and $\delta$ its width. The anchor $X_A$ is the value at the Agilus end and $\Delta_X$ the span to the Vero end, so every function passes through the calibrated value at each end. The sigmoidal shape is chosen to capture the change between the elastomeric and glassy trends: a gradual entry into the transition from the elastomeric plateau, a steep rise, and a smooth exit to the glassy plateau. The asymmetry $\nu$ and the transformation $\sinh^{-1}u$, which is linear near the transition and logarithmic away from it \citep{Johnson1949SystemsOf, Burbidge1988AlternativeTransformations}, together shape that entry and exit. Table~\ref{tab:scaling} lists the parameters of each function, and Figure~\ref{fig:param_scaling} compares them with the calibrated properties of Table~\ref{tab:v2_parameters}.

Increasing Vero content sweeps the glass transition through room temperature (Section~\ref{sec:DMA_Results}), so every property changes most rapidly at the composition where that crossing occurs. The half-rise composition $\phi_V^{50}$, at which $\hat{S}=1/2$ (Table~\ref{tab:scaling}), separates the properties into two groups. The moduli and dimensionless parameters turn over early, between $19$ and $31\%$ Vero, while the mixtures are still elastomeric; $\mu_A$ follows the rubbery-plateau storage modulus measured by DMA. The flow resistances, viscosities, and softening properties turn over later, between $34$ and $56\%$, where room temperature crosses the glass transition of the mixture (Figure~\ref{fig:Fig_DMA_2}c,d) and segmental motion arrests. The resistance of branch~$B$ is the exception, turning over with the moduli, as expected of the slowest branch, which arrests first. The functions also differ in the breadth of their rise. The equilibrium modulus rises over nearly as wide a range of composition as the rubbery plateau it follows, whereas the modulus, exponent, and hardening coefficient of branch~$C$ rise over the narrowest ranges of all. In the elastomeric branches the flow resistances and viscosities rise by more decades than the moduli, capturing each branch to have a greater resistance to flow as Vero content increases. At the Vero end the elastomeric branches cease to flow on the timescale of the experiment and act as elastic springs.

The scaling functions are determined branch by branch. Each is fit to the calibrated properties of the seven compositions in Table~\ref{tab:v2_parameters}, with the Agilus and Vero ends anchored by Sections~\ref{sec:agilus_anchor} and~\ref{sec:vero_anchor}. Branch~$A$ is fixed first, since its locking stretch is shared by the other elastomeric networks, $N_B=N_C=N_A$. The modulus $\mu_A$ is fit to a sigmoid, as motivated by the DMA, and so is $1/N_A$, which varies modestly with composition. Branches~$B$ and~$C$ follow. The modulus $\mu_B$ and the resistance $\tau_{B0}^{\mathrm{ref}}$ have their own functions, as do $\tau_{C0}^{\mathrm{ref}}$, $\alpha_C$, and the linear viscosity $\eta_C^{0}$, which rises by five decades from the Agilus to the Vero end. The modulus $\mu_C$ and the exponent $m_C$ share one function, as the intermolecular mechanism stiffens and becomes more rate sensitive together. Branch~$D$ scaling is determined last. Its modulus and the athermal shear strength $s_0=0.142\,\mu_D$ share one function through their fixed ratio. That ratio sets the elastic shear strain at which the branch reaches its flow resistance, close to the athermal threshold $0.077/(1-\nu_P)$ derived by \citet{Argon1973APolymers}, in which $\nu_P$ is Poisson's ratio. The two softening properties $s_{\mathrm{ss}}/s_0$ and $h_{s}$ share another function, and the linear viscosity $\eta_D^{0}$ has its own. Eleven functions in all carry the composition dependence of the family. For simplicity, the remaining properties are taken to be the same at every composition: the prescribed reference rates $\dot{\gamma}_{B0}^{\mathrm{v}}=\dot{\gamma}_{C0}^{\mathrm{v}}=1~\mathrm{s^{-1}}$, the exponents $m_B=2$ and $q_B=0.2$, the viscosity growth rate $h_{\eta}=80$ and saturation ratio $\eta^{\mathrm{ss}}/\eta^{0}=60$, and the activation energy $\Delta G=29.5~\mathrm{kJ\,mol^{-1}}$ and reference rate $\dot{\gamma}_0^D=1.84~\mathrm{s^{-1}}$ of the glassy flow rule.

Note that given the complexity of the multiple mechanisms transitioning with rate and composition, not every property is uniquely identifiable. Each mechanism is therefore calibrated at the end of the family where it dominates, and the scaling functions supply its properties over the rest of the range, where the response is insensitive to their values. Through these functions the time--composition equivalence of Section~\ref{sec:gD} enters the constitutive description. The properties of each material are set by where its glass transition lies relative to room temperature, a position that composition shifts much as temperature and loading rate do. Those of an untested composition then follow by interpolating the same functions. The model is evaluated against experiment next.

\begin{table}[h!]
\centering
\setlength{\tabcolsep}{2pt}
\caption{Calibrated material properties of the seven compositions used for calibration; A60 is
withheld to validate the model (Section~\ref{sec:validation}). Agilus (A30) and Vero anchor
the family; the properties of the intermediate mixtures follow the composition scaling of
Section~\ref{sec:comp_scaling}. The final column gives the
composition scaling function of each property (Table~\ref{tab:scaling}); a dash marks a property
that is the same at every composition. All Arruda--Boyce springs use the equilibrium locking
stretch, $N_B=N_C=N_A$, and the glassy spring is Neo-Hookean. The flow regularization is
$\xi=10^{-3}$ and the test temperature is $\theta=296$~K.}
\label{tab:v2_parameters}
\vspace{0.5em}
\footnotesize
\begin{tabular}{l*{7}{>{\centering\arraybackslash}p{1.5cm}}>{\centering\arraybackslash}p{1.1cm}>{\centering\arraybackslash}p{1.2cm}}
\toprule
\textbf{Property} & \multicolumn{7}{c}{\textbf{Materials}} & \textbf{Unit} & \textbf{Scaling} \\
\cmidrule(lr){2-8}
 & \textbf{Anchor} & \multicolumn{5}{c}{\textbf{Composition scaling}} & \textbf{Anchor} & & \\
 & \textbf{A30} & \textbf{A50} & \textbf{A70} & \textbf{A85} & \textbf{A95} & \textbf{PP-like} & \textbf{Vero} & & \\
\midrule
\multicolumn{10}{l}{\textbf{Branch~$A$: equilibrium elastic network}} \\
\multicolumn{10}{l}{\quad\emph{Arruda--Boyce spring}} \\
$\mu_A$ & 0.155 & 0.17 & 0.40 & 0.65 & 1.10 & 3.0 & 3.3 & MPa & $f_{\mu_A}$ \\
$N_A$   & 3.0 & 2.95 & 2.6 & 2.5 & 2.45 & 2.12 & 2.1 & -- & $f_{N}$ \\
\midrule
\multicolumn{10}{l}{\textbf{Branch~$B$: elastomeric non-equilibrium (longer-timescale, reptational)}} \\
\multicolumn{10}{l}{\quad\emph{Arruda--Boyce spring} $\;(N_B=N_A)$} \\
$\mu_B$ & 0.12 & 0.121 & 0.18 & 0.522 & 0.75 & 1.28 & 1.3 & MPa & $f_{\mu_B}$ \\
\multicolumn{10}{l}{\quad\emph{Power-law overstress flow}} \\
$\tau_{B0}^{\mathrm{ref}}$ & 0.132 & 0.242 & 1.00 & 8.00 & 22.0 & 338 & 452 & MPa & $f_{\tau_B}$ \\
$\dot{\gamma}_{B0}^{\mathrm{v}}$ & 1.0 & 1.0 & 1.0 & 1.0 & 1.0 & 1.0 & 1.0 & s$^{-1}$ & --- \\
$m_B$ & 2.0 & 2.0 & 2.0 & 2.0 & 2.0 & 2.0 & 2.0 & -- & --- \\
$q_B$ & 0.2 & 0.2 & 0.2 & 0.2 & 0.2 & 0.2 & 0.2 & -- & --- \\
\midrule
\multicolumn{10}{l}{\textbf{Branch~$C$: elastomeric non-equilibrium (shorter-timescale, intermolecular)}} \\
\multicolumn{10}{l}{\quad\emph{Arruda--Boyce spring} $\;(N_C=N_A)$} \\
$\mu_C$ & 1.85 & 1.86 & 2.81 & 4.10 & 5.05 & 5.1 & 5.2 & MPa & $f_{C}$ \\
\multicolumn{10}{l}{\quad\emph{Power-law overstress flow}} \\
$\tau_{C0}^{\mathrm{ref}}$ & 0.262 & 0.263 & 0.510 & 0.900 & 1.33 & 35.0 & 100 & MPa & $f_{\tau_C}$ \\
$\dot{\gamma}_{C0}^{\mathrm{v}}$ & 1.0 & 1.0 & 1.0 & 1.0 & 1.0 & 1.0 & 1.0 & s$^{-1}$ & --- \\
$m_C$ & 1.0 & 1.0 & 2.56 & 4.20 & 4.8 & 5.0 & 5.0 & -- & $f_{C}$ \\
$\alpha_C$ & 6.5 & 6.5 & 14 & 20 & 24 & 24 & 24 & -- & $f_{\alpha_C}$ \\
\multicolumn{10}{l}{\quad\emph{Linear viscous flow}} \\
$\eta_C^0$ & 1.25 & 1.26 & 1.30 & 1.33 & 10.1 & $3.44{\times}10^{4}$ & $1.40{\times}10^{5}$ & MPa\,s & $f_{\eta_C}$ \\
$h_{\eta}$ & 80 & 80 & 80 & 80 & 80 & 80 & 80 & -- & --- \\
$\eta_C^{\mathrm{ss}}/\eta_C^0$ & 60 & 60 & 60 & 60 & 60 & 60 & 60 & -- & --- \\
\midrule
\multicolumn{10}{l}{\textbf{Branch~$D$: glassy non-equilibrium}} \\
\multicolumn{10}{l}{\quad\emph{Neo-Hookean spring}} \\
$\mu_D$ & 0.01 & 0.0242 & 0.495 & 14.5 & 30 & 262 & 340 & MPa & $f_{\mu_D}$ \\
\multicolumn{10}{l}{\quad\emph{Thermally activated flow on athermal strength $s$}} \\
$\Delta G$ & 29.5 & 29.5 & 29.5 & 29.5 & 29.5 & 29.5 & 29.5 & kJ/mol & --- \\
$\dot{\gamma}_{0}^{D}$ & 1.84 & 1.84 & 1.84 & 1.84 & 1.84 & 1.84 & 1.84 & s$^{-1}$ & --- \\
$s_0\;(=0.142\,\mu_D)$ & 0.00142 & 0.00344 & 0.0703 & 2.06 & 4.26 & 37.2 & 48.3 & MPa & $f_{\mu_D}$ \\
$s_{\mathrm{ss}}/s_0$ & 1 & 1 & 1 & 1 & 1 & 0.59 & 0.54 & -- & $f_{s}$ \\
$h_{s}$ & 0 & 0 & 0 & 0 & 0 & 105 & 140 & MPa & $f_{s}$ \\
\multicolumn{10}{l}{\quad\emph{Linear viscous flow}} \\
$\eta_D^0$ & $1.70{\times}10^{-3}$ & $1.35{\times}10^{-2}$ & $1.64{\times}10^{-1}$ & 1.67 & 19.3 & $4.52{\times}10^{4}$ & $1.02{\times}10^{5}$ & MPa\,s & $f_{\eta_D}$ \\
$h_{\eta}$ & 80 & 80 & 80 & 80 & 80 & 80 & 80 & -- & --- \\
$\eta_D^{\mathrm{ss}}/\eta_D^0$ & 60 & 60 & 60 & 60 & 60 & 60 & 60 & -- & --- \\
\bottomrule
\end{tabular}
\end{table}

\begin{table}[h!]
\centering
\caption{Parameters of the composition scaling functions of Eq.~\eqref{eq:scaling_form};
properties listed together share one function. The centers $\phi_V^{0}$ and $\phi_V^{50}$ and the
width $\delta$ are in percent Vero fraction, and the span $\Delta_X$ is the change from the Agilus
to the Vero end, in decades for the properties fitted logarithmically. Entries at $10^{3}$ and
$10^{-8}$ sit at the bounds of the fitting range, where the fitted curve is insensitive to their
value. The two shared functions give $\log_{10}\mu_C=\log_{10}1.85+0.449\,\hat{S}$ and
$m_C=1+4\,\hat{S}$ ($R^2=0.9992$ and $0.9995$), and $s_{\mathrm{ss}}/s_0=1-0.46\,\hat{S}$ and
$h_{s}=140\,\hat{S}$~MPa ($R^2=0.9958$).}
\label{tab:scaling}
\vspace{0.5em}
\small
\begin{tabular}{lllccccccc}
\toprule
\textbf{Function} & \textbf{Applies to} & \textbf{Scale} & $k$ & $\delta$ (\%) & $\phi_V^{0}$ (\%) & $\nu$ & $\phi_V^{50}$ (\%) & $\Delta_X$ & $R^2$ \\
\midrule
\multicolumn{10}{l}{\emph{Branch A}} \\
$f_{\mu_A}$ & $\mu_A$ & logarithmic & 64.7 & $10^{3}$ & 21.6 & 0.011 & 27.5 & $+1.33$ & 0.9948 \\
$f_{N}$ & $1/N_A$ & linear & 186 & $10^{3}$ & 25.3 & $10^{3}$ & 31.3 & $+0.143$ & 0.9798 \\
\midrule
\multicolumn{10}{l}{\emph{Branch B}} \\
$f_{\mu_B}$ & $\mu_B$ & logarithmic & 0.560 & 3.33 & 22.1 & $10^{-8}$ & 23.5 & $+1.03$ & 0.9956 \\
$f_{\tau_B}$ & $\tau_{B0}^{\mathrm{ref}}$ & logarithmic & 67.3 & $10^{3}$ & 22.4 & 0.034 & 27.9 & $+3.53$ & 0.9931 \\
\midrule
\multicolumn{10}{l}{\emph{Branch C}} \\
$f_{C}$ & $\mu_C$, $m_C$ & see caption & 189 & $10^{3}$ & 17.8 & 0.016 & 19.7 & see caption & see caption \\
$f_{\tau_C}$ & $\tau_{C0}^{\mathrm{ref}}$ & logarithmic & 105 & $10^{3}$ & 52.3 & 2.61 & 48.2 & $+2.58$ & 0.9953 \\
$f_{\alpha_C}$ & $\alpha_C$ & linear & 190 & $10^{3}$ & 17.8 & 0.21 & 19.4 & $+17.5$ & 0.9988 \\
$f_{\eta_C}$ & $\eta_C^0$ & logarithmic & 88.5 & $10^{3}$ & 43.3 & 0.082 & 47.0 & $+5.05$ & 0.9999 \\
\midrule
\multicolumn{10}{l}{\emph{Branch D}} \\
$f_{\mu_D}$ & $\mu_D$, $s_0$ & logarithmic & 1.49 & 2.29 & 18.7 & $10^{3}$ & 20.2 & $+4.53$ & 0.9968 \\
$f_{\eta_D}$ & $\eta_D^0$ & logarithmic & 251 & $10^{3}$ & 47.1 & 114 & 34.2 & $+7.78$ & 0.9995 \\
$f_{s}$ & $s_{\mathrm{ss}}/s_0$, $h_{s}$ & linear & 122 & $10^{3}$ & 53.0 & $10^{-8}$ & 56.0 & see caption & 0.9958 \\
\bottomrule
\end{tabular}
\end{table}

\section{Model evaluation across compositions and rates}
\label{sec:Model_Exp}

The Agilus and Vero anchors were calibrated in Section~\ref{sec:calibration}. This section compares the model with experiment across the five intermediate mixtures, whose properties follow the composition scaling of Section~\ref{sec:comp_scaling}, resolves the deformation and the stress carried by each branch in a mixture that lies within the transition, and then maps how the mechanisms share the stress, and the work they store and dissipate, across the entire family.

\begin{figure}[htbp]
    \centering
    \includegraphics[width=\textwidth]{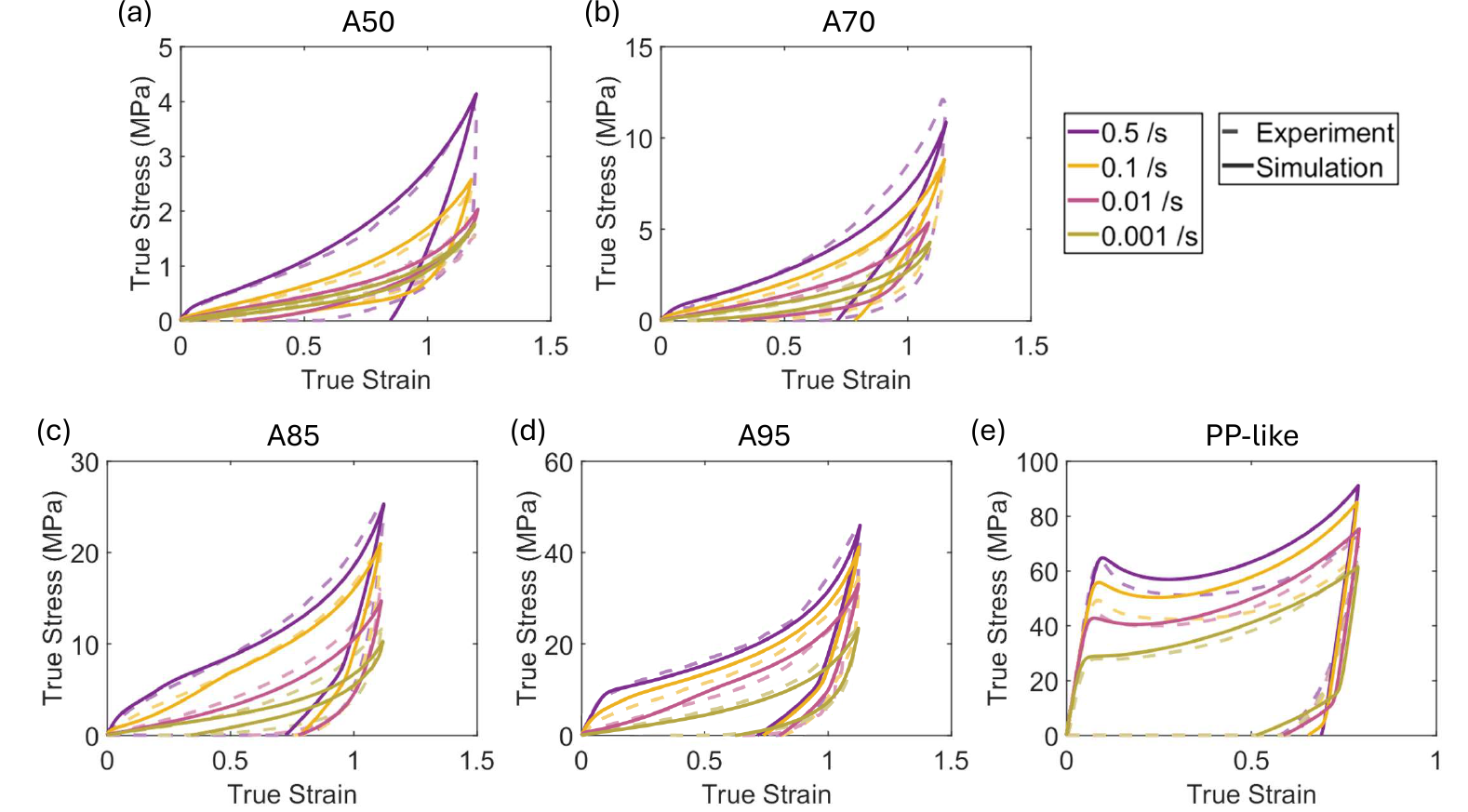}
    \caption{Model versus experiment for the intermediate digital mixtures spanning the rubbery-to-glassy transition: (a)~A50, (b)~A70, (c)~A85, (d)~A95, and (e)~PP-like. The material properties follow the composition scaling of Section~\ref{sec:calibration} (Table~\ref{tab:v2_parameters}). Each panel shows the total true stress versus true strain load--unload response in uniaxial compression across strain rates (dashed, experiment; solid, model).}
    \label{fig:mixtures_fit}
\end{figure}

\begin{figure}[htbp]
    \centering
    \includegraphics[width=\textwidth]{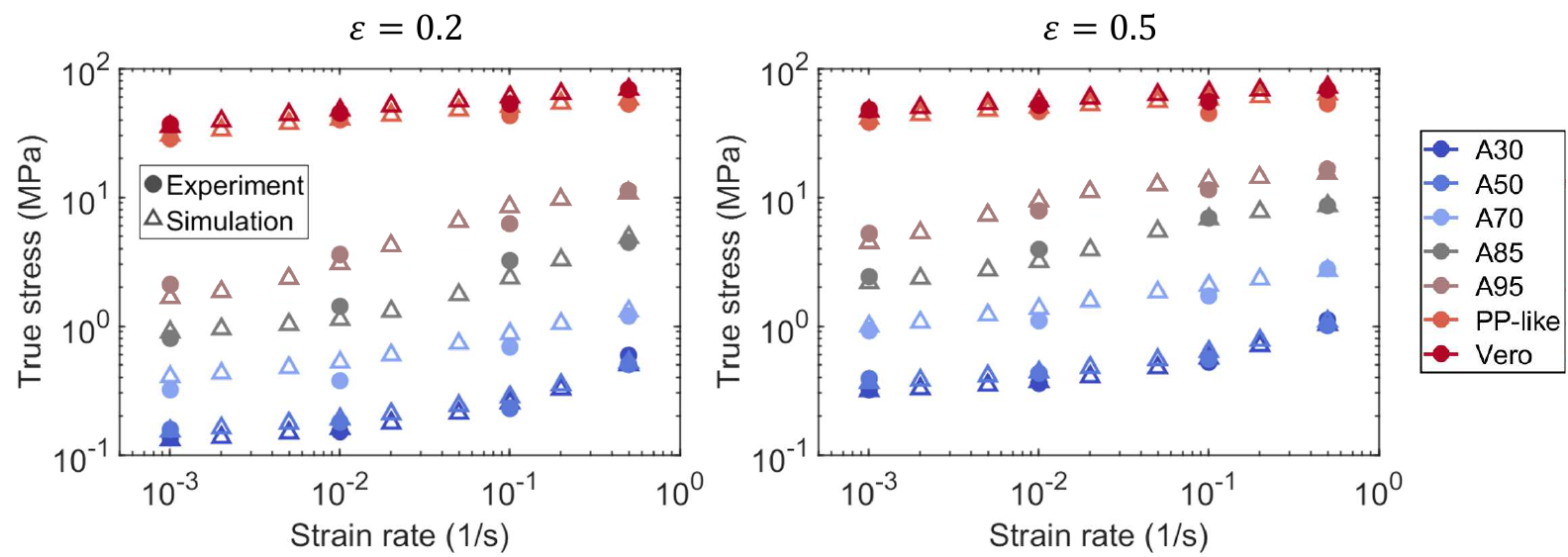}
    \caption{True stress at true strains of $0.2$ and $0.5$ versus strain rate, model versus experiment. Filled circles, the measurements of Figure~\ref{fig:Zwick_modulus}(c,~d); open triangles, the model evaluated at nine rates from $0.001$ to $0.5~\mathrm{s^{-1}}$, adding five untested rates. The denser rates resolve the transition between the low-rate and high-rate sensitivities that the four tested rates bracket; the transition shifts to lower rates with increasing Vero content, as in Figure~\ref{fig:Zwick_modulus}.}
    \label{fig:stress_rate_model}
\end{figure}

\begin{figure}[htbp]
    \centering
    \includegraphics[width=\textwidth]{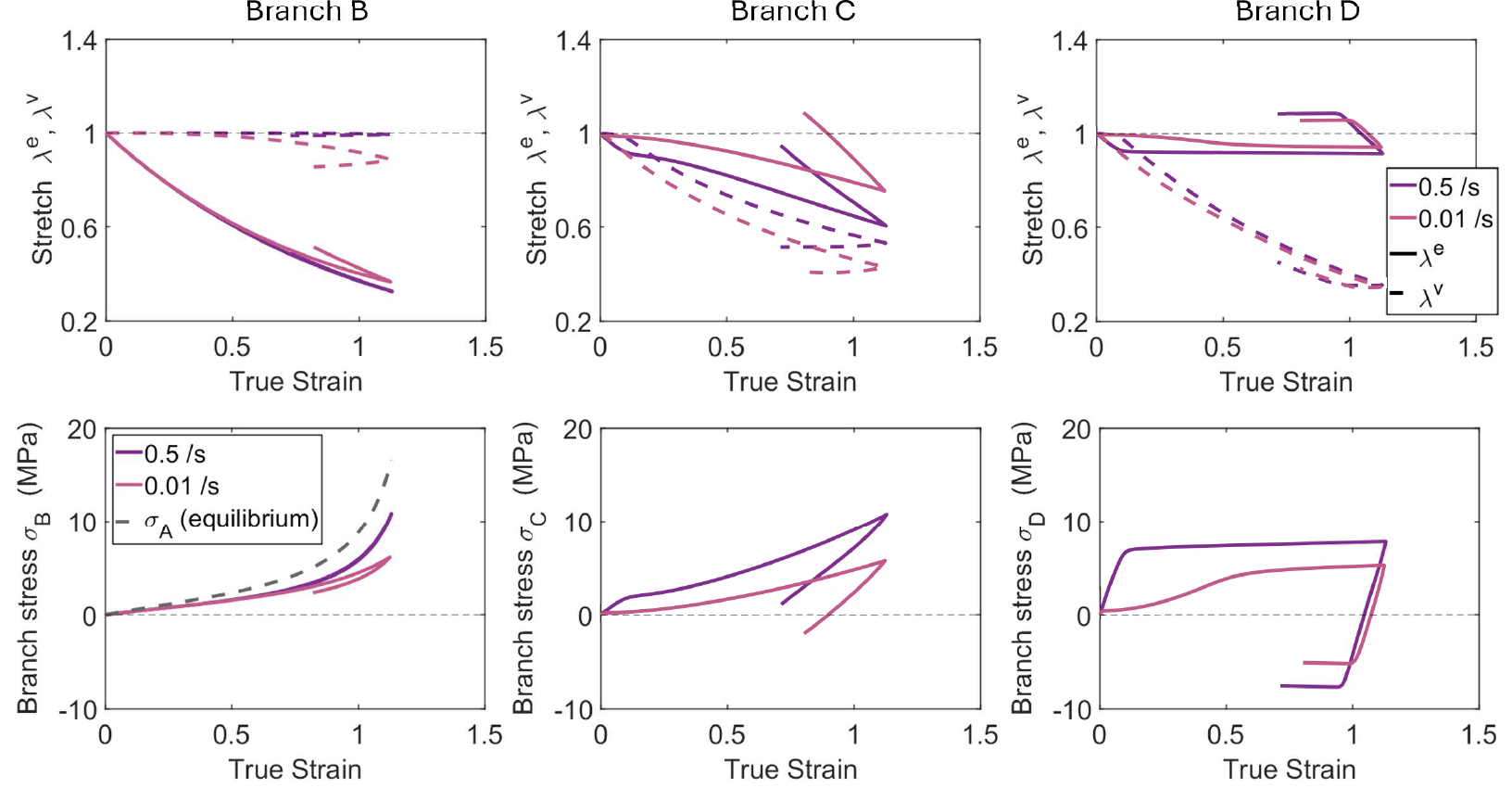}
    \caption{Branch-level decomposition of the A95 response at $0.5$ and $0.01~\mathrm{s^{-1}}$. The columns are branches~$B$, $C$ and~$D$, and both rows are plotted against the applied true strain. Top row: elastic and viscous stretches $\lambda_i^{\mathrm{e}}$ (solid) and $\lambda_i^{\mathrm{v}}$ (dashed), with the line at unity marking the undeformed state, below it compression and above it tension. Bottom row: the branch stress $\sigma_i$, positive in compression; the grey dashed curve in the branch-$B$ panel is the equilibrium stress $\sigma_A$, which is rate independent and free of hysteresis, shown for scale. Purple, $0.5~\mathrm{s^{-1}}$; magenta, $0.01~\mathrm{s^{-1}}$; each curve is traced to the strain at which the total stress returns to zero.}
    \label{fig:lambdaev_sigma_A95_fit}
\end{figure}

\subsection{Intermediate mixtures across the transition}
\label{sec:model_vs_exp_intermediate}

As the Vero fraction rises, the effective glass transition crosses room temperature and the response turns from elastomeric to glassy (Figure~\ref{fig:Fig_DMA_2}). Figure~\ref{fig:mixtures_fit} compares model and experiment for the five intermediate mixtures at all four rates, with the material properties taken from the composition scaling of Section~\ref{sec:comp_scaling}. Peak stress spans from under $2~\mathrm{MPa}$ at A50 and the lowest rate to about $90~\mathrm{MPa}$ at PP-like and the highest, and the character of the response changes across that range. A50 and A70 stay largely recoverable, with rate-dependent hysteresis and modest residual strain; by A95 a clear rollover has developed and the hysteresis loops have increased in area; at PP-like the response is fully glassy, with yield, pronounced hysteresis, and a large residual strain at every rate. In the model this progression comes from the modulus and flow resistance of branch~$C$ increasing as the composition traverses the transition from the rubbery regime, and from the glassy branch~$D$ engaging as the compositions approach the glassy regime. At each composition the model follows the initial slope and its rate dependence, the rollover, the width of the load--unload loop, the residual strain, and the upturn at large strain, simultaneously at all four rates. Figure~\ref{fig:stress_rate_model} replots the true stress at $\varepsilon=0.2$ and $0.5$ against strain rate, with the model evaluated at nine rates, five of them untested. The model follows the measurements at the four tested rates and, between them, traces a smooth rise in rate sensitivity whose onset moves to lower rates as Vero content increases, the same composition-dependent shift seen in the data of Figure~\ref{fig:Zwick_modulus}. The rise is largest for the mixtures within the transition and nearly flat for PP-like and Vero, which stay glassy across the window.

We note two inconsistencies between the model and the measurements. First, at the higher rates the measured stress decreases almost vertically at load reversal while the model exhibits a steep unloading slope transitioning to nonlinear reverse flow; we suspect a machine response time governs this near-vertical drop, but could not confirm it. Second, for PP-like, which is glassy over the tested rates, the isothermal model overpredicts the large-strain hardening at the two fastest rates, since the measured response shows thermal softening, as also seen at the Vero anchor (Section~\ref{sec:vero_anchor}).

\subsection{Branch-level mechanism of a transitional mixture}
\label{sec:model_vs_exp_mechanism}

To identify how each mechanism contributes to the rate-dependent stress--strain behavior, and how the elastic and viscous deformation of each branch evolves, we resolve the A95 response of Figure~\ref{fig:mixtures_fit}d. A95 lies near the middle of the room-temperature glass transition (Figure~\ref{fig:Fig_DMA_2}), where the elastomeric and the glassy mechanisms are both active, so the partition is most fully developed there. Figure~\ref{fig:lambdaev_sigma_A95_fit} shows the three non-equilibrium branches at a higher rate ($0.5~\mathrm{s^{-1}}$) and a lower rate ($0.01~\mathrm{s^{-1}}$): the elastic and viscous stretches $\lambda_i^{\mathrm{e}}$ and $\lambda_i^{\mathrm{v}}$ (top row) and the stresses $\sigma_i$ (bottom row), which together with the equilibrium network~$A$ sum to the total of Figure~\ref{fig:mixtures_fit}d.

Each branch accommodates the applied deformation by both its elastic and viscous stretches, and this accommodation depends on the strain rate. At $0.5~\mathrm{s^{-1}}$ branch~$B$ has insufficient time to relax and hence does not flow: $\lambda_B^{\mathrm{v}}$ stays at unity while $\lambda_B^{\mathrm{e}}$ follows the applied stretch down to $0.33$. The branch~$C$ deformation is nearly equally carried by elastic and viscous mechanisms, reaching $\lambda_C^{\mathrm{e}}=0.61$ and $\lambda_C^{\mathrm{v}}=0.53$, consistent with its shorter relaxation time. The glassy branch~$D$ flows upon reaching its critical stress. At $0.01~\mathrm{s^{-1}}$ branch~$B$, the slowest mechanism, still accommodates most of its deformation elastically, while branches~$C$ and~$D$ accommodate most of their deformation by viscous flow. On unloading the elastic stretches recover and in some cases pass through unity into tension, which is balanced by the compression still carried by the other branches. The tension drives a reverse viscous flow, as seen in the evolution of $\lambda_i^{\mathrm{v}}$. These interchanges between the branches govern the hysteresis loops and the recovery.

The partition of the stress across the mechanisms also accounts for the wide spread of the A95 curves with rate in Figure~\ref{fig:mixtures_fit}d. At $0.5~\mathrm{s^{-1}}$ all three non-equilibrium branches are active in substantive but different ways as the strain increases, whereas at $0.01~\mathrm{s^{-1}}$ the overstress of branch~$C$ has partly relaxed and the rate-independent equilibrium network carries nearly half the peak stress. Strain rate here acts as composition does, carrying one material over the rubbery-to-glassy range that rising Vero content produces across the family. One constitutive structure, with composition entering only through the scaling laws of Section~\ref{sec:calibration}, thus reproduces the nonlinear, rate-dependent large-deformation response of the entire family, from elastomeric hysteresis to glassy yield. The partition is examined across all compositions and rates next.

\subsection{Deformation mechanisms across the rubbery-to-glassy transition}
\label{sec:mechanism}
Having evaluated the model against experiment, we now use it to identify which mechanism carries the load across the family, and how the work done divides between storage and dissipation. The branches contribute additively to the stress, so the fraction carried by each measures the role of its mechanism at any composition, rate and strain; resolving the energy in the same way separates storage from dissipation.

\begin{figure}[htbp]
    \centering
    \includegraphics[width=0.9\textwidth]{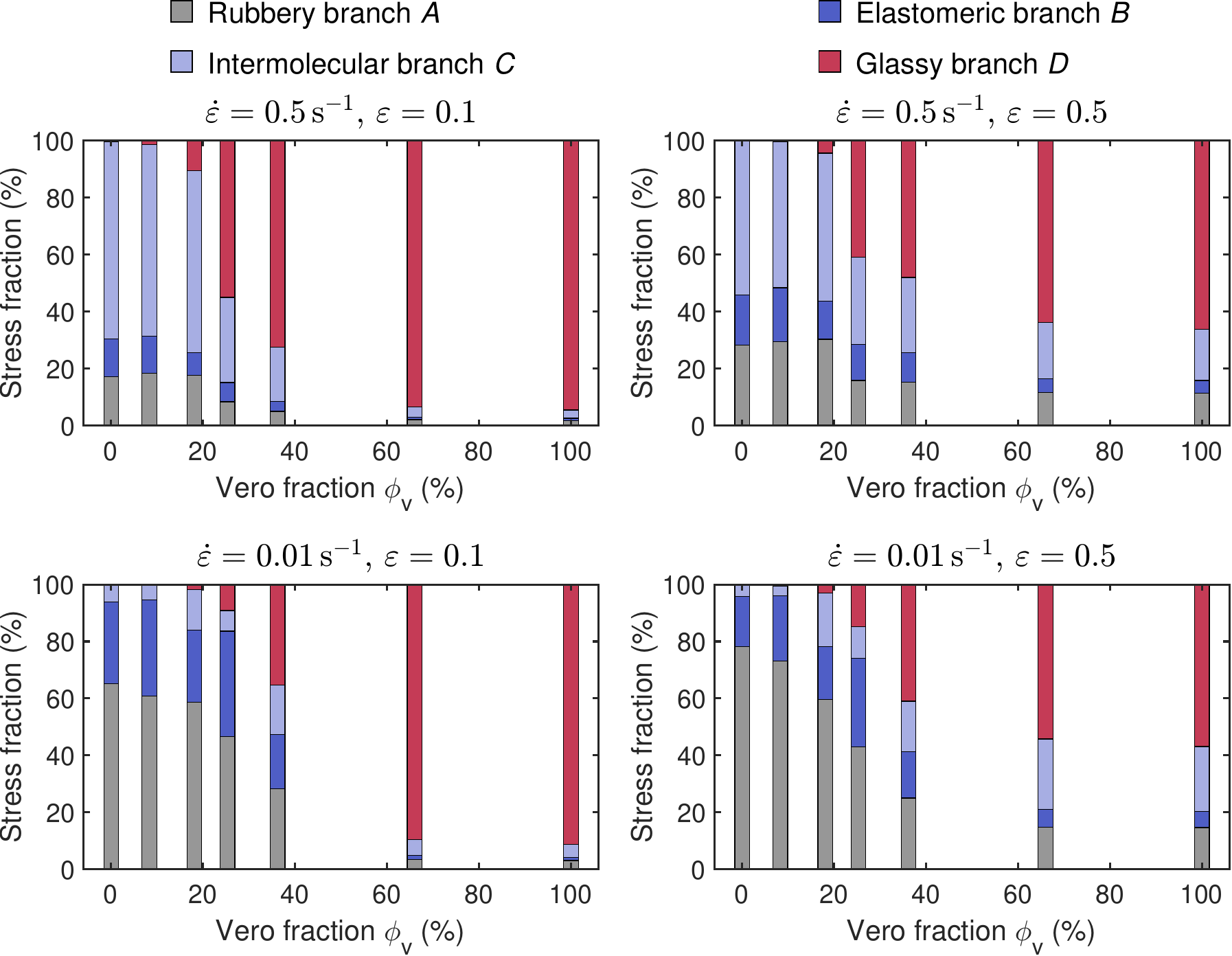}
    \caption{Model-predicted partition of the uniaxial-compression stress among the constitutive mechanisms: the rubbery branch~$A$ ($\sigma_A$, grey), the elastomeric branch~$B$ ($\sigma_B$, dark blue), the intermolecular branch~$C$ ($\sigma_C$, light blue), and the glassy branch~$D$ ($\sigma_D$, red), each as a percentage of the total Cauchy stress at the seven calibrated compositions. Panels span two loading rates ($\dot\varepsilon=0.5$ and $0.01~\mathrm{s^{-1}}$) and two strain levels ($\varepsilon=0.1$ and $0.5$). The load passes from the elastomeric to the glassy mechanism as $\phi_V$ rises; higher rate and lower strain both enlarge the glassy fraction.}
    \label{fig:stress_partition}
\end{figure}

Figure~\ref{fig:stress_partition} resolves the uniaxial compression stress into its four branch contributions, the equilibrium network $\sigma_A$, the elastomeric non-equilibrium branches $\sigma_B$ and $\sigma_C$, and the glassy branch $\sigma_D$, at two rates and two levels of strain for each composition. At the Agilus end the glassy branch is negligible, and the load is shared between the equilibrium network and the elastomeric non-equilibrium branches: at A30 and $\varepsilon=0.5$ the equilibrium share is $78\%$ at $0.01~\mathrm{s^{-1}}$ but $28\%$ at $0.5~\mathrm{s^{-1}}$, so the equilibrium network dominates at the lower rate and branches~$B$ and~$C$ dominate at the higher rate. At the Vero end the glassy branch dominates at both rates, with a growing elastomeric share at large stretch, where chain orientation stiffens the network: at $0.5~\mathrm{s^{-1}}$ that share rises from $6\%$ at $\varepsilon=0.1$ to $34\%$ at $\varepsilon=0.5$. Across the composition range the elastomeric non-equilibrium mechanisms, which carry the rate dependence at the Agilus end, give way to the glassy one: at $0.5~\mathrm{s^{-1}}$ and $\varepsilon=0.5$ their combined share falls from $72\%$ at A30 to $22\%$ at Vero, while the glassy share rises from below $1\%$ to $66\%$. Within the transition, raising the rate enlarges the glassy fraction, since branch~$D$ engages when the loading is fast relative to its relaxation time.

Resolving the two elastomeric branches separates their timescales. At $0.5~\mathrm{s^{-1}}$ the shorter-timescale branch~$C$ carries most of the non-equilibrium stress, roughly three to four times branch~$B$ at $\varepsilon=0.5$ in the Agilus-rich materials. At $0.01~\mathrm{s^{-1}}$ that ordering reverses, because branch~$C$ has relaxed while branch~$B$ has not: at A30 and $\varepsilon=0.5$ branch~$B$ holds about $18\%$ at both rates while the share of branch~$C$ falls from $54$ to $4\%$. In the Vero-rich materials both branches are arrested at both rates, so the reversal disappears and their shares change little with rate. The rubbery-to-glassy progression seen in the experiments thus appears in the model as a continuous shift in which mechanism carries the load, not as a change of constitutive form.

\begin{figure}[htbp]
    \centering
    \includegraphics[width=\textwidth]{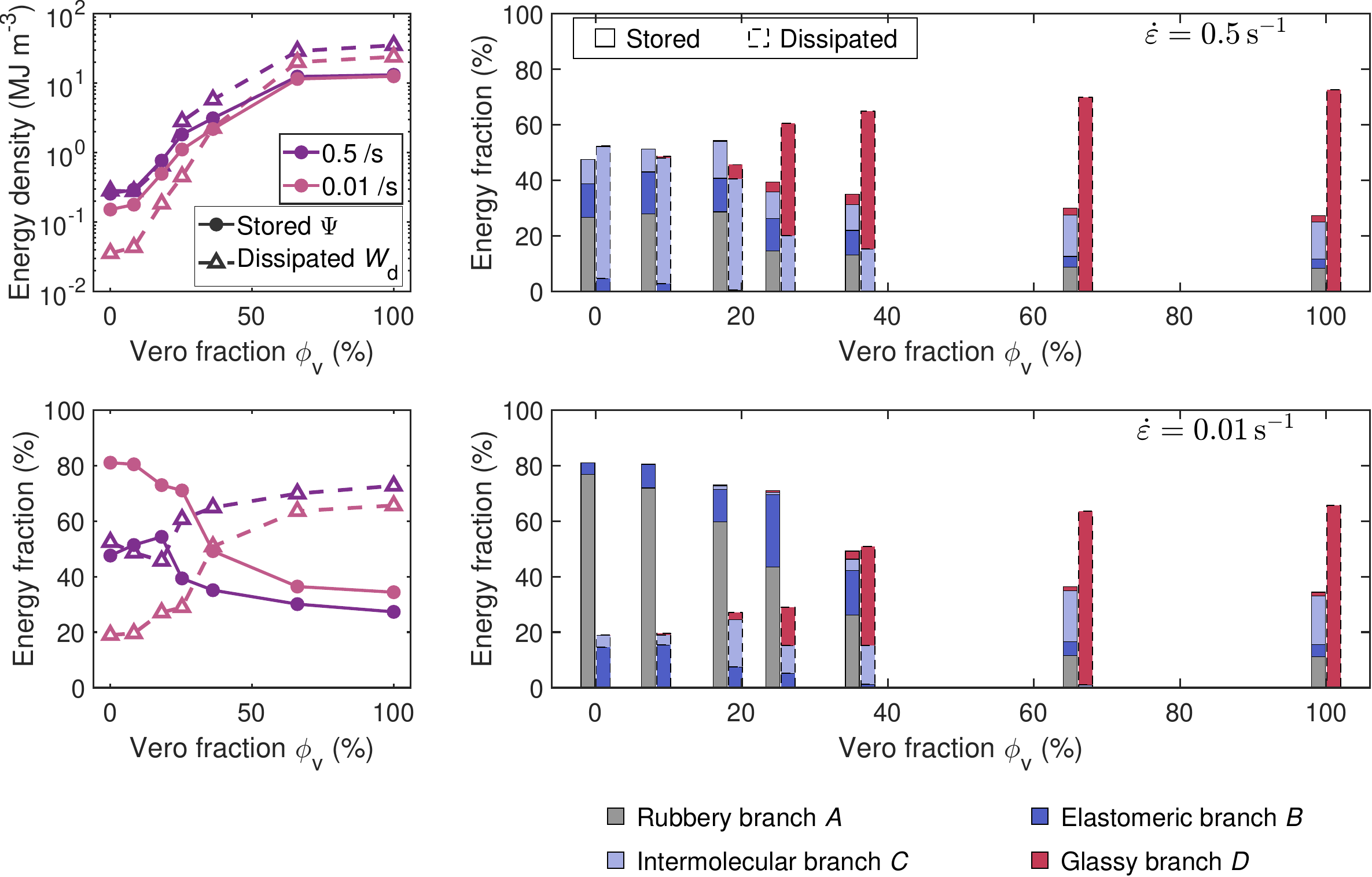}
    \caption{Model-predicted partition of the input work into stored and dissipated energy, Eq.~\eqref{eq:work_split}, at the common strain $\varepsilon=0.7$, for the seven calibrated compositions. Left: the free energy $\Psi$ stored in the springs (solid, filled circles) and the energy $W_d$ dissipated by the dashpots (dashed, open triangles) at the two rates ($\dot\varepsilon=0.5$ and $0.01~\mathrm{s^{-1}}$), as densities above and as fractions of the input work $W=\Psi+W_d$ below. Right: the same two fractions resolved by branch, the left bar of each pair the stored energy over branches~$A$ to~$D$ (solid outline) and the right bar the dissipated energy over~$B$, $C$ and~$D$ (dashed outline). Every segment is a percentage of $W$, so each pair sums to $100\%$.}
    \label{fig:energy_partition}
\end{figure}

Along the loading path the work done per unit reference volume splits into the free energy $\Psi$ stored in the four springs, Eq.~\eqref{eq:Psi_total}, and the energy $W_d$ dissipated by the three dashpots, the time integral of Eq.~\eqref{eq:Dissipation_closed}, with $W=\Psi+W_d$ by Eq.~\eqref{eq:work_split}. Figure~\ref{fig:energy_partition} evaluates both at a common strain $\varepsilon=0.7$, as energy densities and as fractions of $W$ on the left, and resolved by branch on the right. Both energies rise by about two decades with Vero content. The share of stored energy is largest at low rate and low Vero content, $81\%$ of the work for A30 at $0.01~\mathrm{s^{-1}}$, and falls as composition or rate rises, reaching $27\%$ for Vero at $0.5~\mathrm{s^{-1}}$. For A30 the dissipated share rises from $19$ to $52\%$ between the two rates.

Carrying the load and storing or dissipating the work are distinct roles, and the branch-resolved bars show their division. At the Agilus end the stored energy resides in the equilibrium network, $77\%$ of the input work at A30 and $0.01~\mathrm{s^{-1}}$. At the Vero end it resides mainly in branches~$B$ and~$C$, which are arrested and act as elastic springs; the glassy branch yields at small elastic stretch and stores only $2\%$ of the work at $0.5~\mathrm{s^{-1}}$. The dissipation moves in the opposite direction, from the elastomeric branches to the glassy one: at $0.5~\mathrm{s^{-1}}$ branch~$C$ dissipates $47\%$ of the work at A30 and $20\%$ at A85 as its flow resistance rises, while branch~$D$ climbs from $5\%$ at A70 to $40\%$ at A85 and $73\%$ at Vero. In the Agilus-rich mixtures the dissipating branch depends on rate, branch~$C$ at $0.5~\mathrm{s^{-1}}$ and branch~$B$ at $0.01~\mathrm{s^{-1}}$, the same exchange of timescales seen in the stress. At the higher rate the stored share peaks at A70, in the handover between the two dissipating mechanisms, where branch~$C$ has slowed and branch~$D$ has yet to engage. This energy partition carries a direct implication for material design: composition tunes a printed material from mainly storing the input work to mainly dissipating it. The balance of storage and dissipation can therefore be tailored, in a single mixture or region by region across a graded multi-material structure, to the operating rate and strain.

\section{Predictive test of the composition scaling}
\label{sec:validation}

\begin{figure}[htbp]
    \centering
    \includegraphics[width=0.45\textwidth]{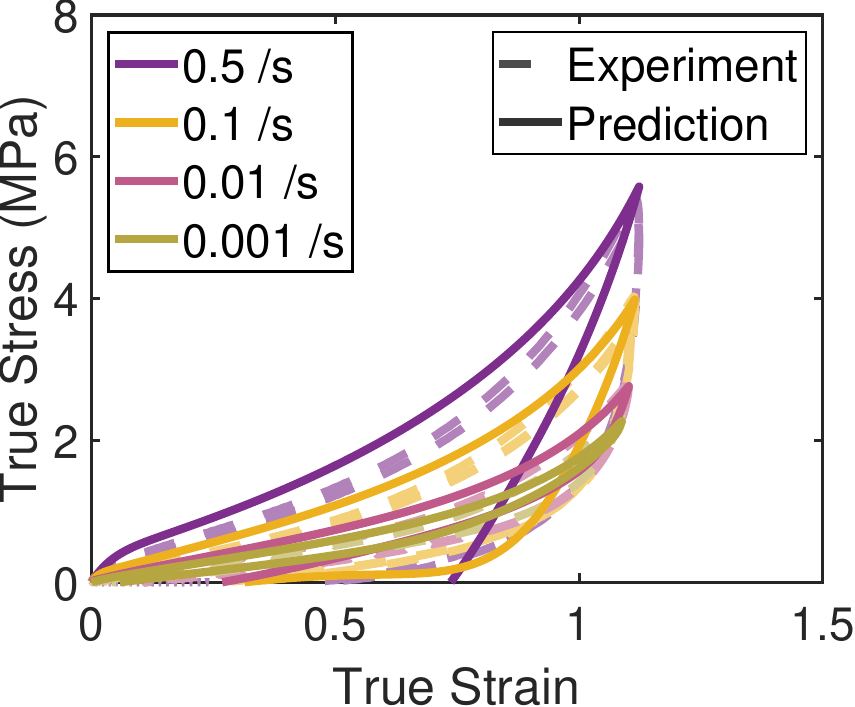}
    \caption{Predicted and measured uniaxial compression response of A60 ($\phi_V=13.2\%$) at four strain rates. Solid curves, the model with the properties of Table~\ref{tab:A60_predicted}; dashed curves, the measured load--unload cycles, every repeat overlaid. A60 was withheld from the calibration, so the comparison is a blind prediction.}
    \label{fig:A60_validation}
\end{figure}

Assuming composition enters the model only through scaling of the material properties, here we evaluate a remaining Stratasys digital mixture that was not used in the calibration. A60, at $\phi_V=13.2\%$, was printed and tested. We note that this composition lies between A50 ($\phi_V=8.3\%$) and A70 ($\phi_V=18.1\%$). Also note that A70 has entered the transition regime. The A60 model properties, obtained via the scaling laws, are given in Table~\ref{tab:A60_predicted}. Figure~\ref{fig:A60_validation} compares the prediction with the measurement at four rates spanning nearly three decades. The prediction captures the experimental response reasonably well: the nonlinear upturn of each rate together with the rate dependence of the initial slope, rollover, tangential stiffness and hysteresis. We note that the prediction slightly overshoots the stress at the highest rate, $0.5~\mathrm{s^{-1}}$.

\begin{table}[h!]
\centering
\setlength{\tabcolsep}{4pt}
\caption{Properties of A60 ($\phi_V=13.2\%$) determined by the composition scaling functions of
Table~\ref{tab:scaling}. The remaining properties are
composition independent and take the values of Table~\ref{tab:v2_parameters}.}
\label{tab:A60_predicted}
\vspace{0.5em}
\small
\begin{tabular}{cc@{\hskip 14pt}cc@{\hskip 14pt}ccccc@{\hskip 14pt}ccc}
\toprule
\multicolumn{2}{c}{Branch~$A$} & \multicolumn{2}{c}{Branch~$B$} &
\multicolumn{5}{c}{Branch~$C$} & \multicolumn{3}{c}{Branch~$D$} \\
\cmidrule(r){1-2}\cmidrule(r){3-4}\cmidrule(r){5-9}\cmidrule(r){10-12}
$\mu_A$ & $N_A$ & $\mu_B$ & $\tau_{B0}^{\mathrm{ref}}$ &
$\mu_C$ & $\tau_{C0}^{\mathrm{ref}}$ & $m_C$ & $\alpha_C$ & $\eta_C^{0}$ &
$\mu_D$ & $s_0$ & $\eta_D^{0}$ \\
(MPa) & (--) & (MPa) & (MPa) & (MPa) & (MPa) & (--) & (--) & (MPa\,s) & (MPa) & (MPa) & (MPa\,s) \\
\midrule
0.259 & 2.75 & 0.141 & 0.462 & 2.05 & 0.385 & 1.39 & 8.97 & 1.25 & 0.0615 & 0.00873 & 0.0496 \\
\bottomrule
\end{tabular}
\end{table}

The accuracy of such a prediction depends on where the mixture sits in the composition range. A60 lies between the elastomeric A50 and A70, which has entered the transition. Therefore, it does provide a modest test of whether the predictive scaling captures how the rate dependence changes with composition in this regime. Ideally, a digital mixture more centrally in the steep portion of the transition would be a greater test. There the glass transition sweeps through room temperature, and a few percent of Vero fraction moves the shear moduli and flow resistances by close to an order of magnitude. Unfortunately, additional digital mixes are not available; future work will evaluate more discrete digital mixing at small scales for further assessment of the model. At present, the compositional scaling provides an opportunity to assess the features across a wider range of $\phi_V$. The predictive nature of the model now enables simulation-based design of structured, architected, and graded materials and products for nonlinear rate-dependent performance, energy storage and dissipation.

\section{Predictive test across states of deformation}
\label{sec:states}

The material properties were determined from uniaxial compression alone (Section~\ref{sec:calibration}). With these properties, the model is used here to predict two other states of deformation, plane-strain compression and uniaxial tension, for three representative compositions, A30, A70 and A95, which range from the elastomeric response to the middle of the transition. The three states of deformation act to evolve the molecular network orientation differently with large strain and are considered tests of elastomeric and glassy polymer constitutive models. Uniaxial compression orients the network towards an equibiaxial orientation, in the plane normal to the compression axis; uniaxial tension orients a network towards a uniaxial orientation along the loading axis; plane-strain compression provides a network orientation that extends primarily along the free direction, respecting the constraint direction. The model accounts for these differences through the eight-chain networks of branches~$A$, $B$ and~$C$, whose chain stretch and orientation depend on the state of deformation.

\begin{figure}[!htb]
    \centering
    \includegraphics[width=\textwidth]{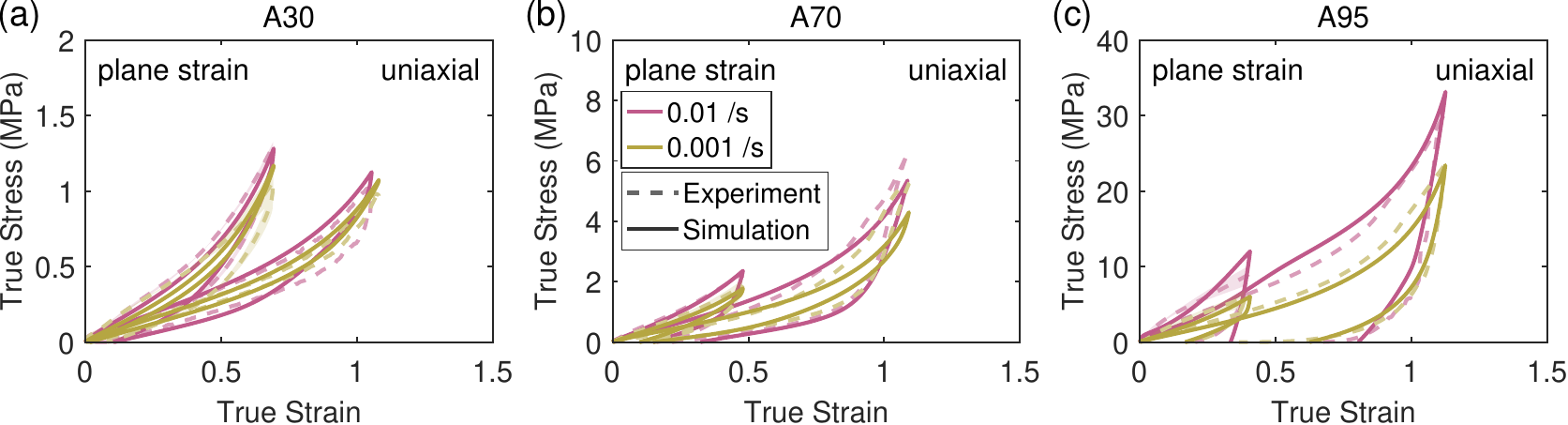}
    \caption{Predicted and measured plane-strain and uniaxial compression of (a)~A30, (b)~A70 and (c)~A95 at $0.01$ and $0.001~\mathrm{s^{-1}}$. Dashed curves, experiment; solid curves, model with the properties from uniaxial compression. The plane-strain experiment is the mean of the tests, with their range shaded; the uniaxial experiment is the calibration test.}
    \label{fig:planestrain_uniaxial}
\end{figure}

Figure~\ref{fig:planestrain_uniaxial} compares plane-strain and uniaxial compression at strain rates of $0.01$ and $0.001~\mathrm{s^{-1}}$. The model captures the higher stress in plane strain compared to that in uniaxial compression for all three compositions at both rates, together with the rate dependence and hysteresis of the plane-strain response. The measured plane-strain stress is slightly higher than that predicted, which may partly reflect friction at the walls of the channel die, not included in the model. A comparison of uniaxial tension to uniaxial compression is provided in the appendix (Figure~\ref{fig:tension_appendix}). The specimens printed in three in-plane orientations show a nearly isotropic response, with differences between orientations comparable to the variability between specimens. The model captures the state-of-deformation dependence of each material well. A30 is well predicted; A70 and A95 predictions fall within the variability of the measurements (we also note that the maximum strain in A70 and A95 was limited due to their respective tensile failure strain).  The model, calibrated in one state of deformation, thus captures the main features of all three deformation states, together with the dependence of the response on composition and strain rate.

\FloatBarrier
\section{Conclusions and future work}
\label{sec:conclusion}

The composition-dependent nonlinear viscoelastic--viscoplastic behavior of digitally mixed Agilus--Vero polymers was characterized and modeled across the full composition range. Small-strain DMA and large-deformation uniaxial compression over nearly three orders of magnitude in strain rate document that the response evolves continuously with composition and rate: from recoverable elastomeric hysteresis, through the leathery transition, to glassy yield with post-yield softening, hardening, and substantial residual strain. DMA shows that each digital mixture has a single glass transition that shifts to higher temperature with both Vero fraction and rate. Composition therefore acts on the position of the transition in an intrinsic material manner analogous to how temperature and loading rate act as extrinsic loading conditions. Hence, a time--composition equivalence places every mixture, tested or not, on one rubbery-to-glassy axis.

This equivalence motivates describing the family of mixtures with one finite-strain constitutive structure. An equilibrium hyperelastic network acts in parallel with three non-equilibrium branches representing the reptational, intermolecular, and glassy mechanisms, and the branches contribute additively to the stress. Composition enters only through the material properties, anchored at the two endpoints, elastomeric Agilus and glassy Vero, and interpolated by smooth, physically motivated scaling laws. Rate dependence follows from the flow-rule kinetics. The calibrated properties of the seven materials collapse onto these laws, so the full range of behavior follows from one constitutive structure with a few smoothly varying properties. The model reproduces the experimental compression response of all seven compositions over the full rate range. The description extends to untested mixtures: the withheld A60 composition was predicted from the scaling laws alone. It also extends to other states of deformation: the properties determined from uniaxial compression predict the plane-strain compression and uniaxial tension responses of three representative compositions.

The model also identifies which mechanism carries the load, and how the work done divides between storage and dissipation. Resolving the predicted stress into its branches shows the load shifting from the elastomeric to the glassy mechanism as Vero content and rate rise. Strain rate acts on the mechanism as composition does. Resolving the work the same way shows the stored share largest at low rate and low Vero content, falling as either rises. Dissipation mechanisms change in nature and magnitude from the elastomeric domain to the glassy domain. Composition is the single design variable. The framework thus supports the design of structured and architected materials and products with nonlinear rate-dependent performance, energy storage and dissipation.

Several extensions follow naturally. The present formulation is isothermal; temperature-dependent behavior together with thermo-mechanical coupling, calibrated against large-deformation experiments over a range of temperature, would extend the time--temperature--composition equivalence established for the linear viscoelastic response~\citep{ShenBoyceTCS} to finite deformation. Extending the model beyond quasi-static rates to impact loading, where even Agilus may become leathery or glassy, would activate the glassy branch~$D$ in Agilus, which would then also contribute to the scaling across composition. Furthermore, high rates also result in deformation-induced heating and thermal softening. The constitutive framework should also guide modeling of other digitally mixed families, for example systems whose constituents differ in crosslink density, which would establish the generality of the composition scaling beyond the Agilus--Vero pair. The physically interpretable property set also suits data-driven extension, learning smooth relations among composition, temperature, rate, and the material properties to interpolate across the design space; a companion study takes a first step in this direction, discovering composition-dependent hyperelastic and viscoelastic constitutive models for the A30--A95 mixtures directly from experimental data~\citep{GarciaAvilaDataDriven}. Finally, finite-element implementation and application to spatially programmed multi-material structures, currently underway, will enable predictive design of rate-dependent, tunable mechanical response in voxel-printed architectures.

\section*{CRediT authorship contribution statement}
\textbf{Beijun Shen}: Conceptualization, Methodology, Investigation, Formal analysis, Data curation, Software, Validation, Visualization, Writing -- original draft, Writing -- review \& editing.
\textbf{Mary C.\ Boyce}: Conceptualization, Methodology, Supervision, Funding acquisition, Resources, Writing -- review \& editing.

\section*{Declaration of competing interest}
The authors declare that they have no known competing financial interests or personal relationships that could have appeared to influence the work reported in this paper.

\section*{Acknowledgments}
This work was supported by Columbia University. The authors thank Prof.\ Yevgeniy Yesilevskiy for laboratory management and equipment maintenance, and Dr.\ Will Hunnicutt for DMA training and coordinating equipment use at the Carleton Laboratory at Columbia University. We also acknowledge group discussions with Prof.\ Adrian Buganza Tepole and Mr.\ Josue Garcia Avila.

\appendix
\setcounter{figure}{0}

\section{Uniaxial tension}
\label{app:tension}

\begin{figure}[htbp]
    \centering
    \includegraphics[width=\textwidth]{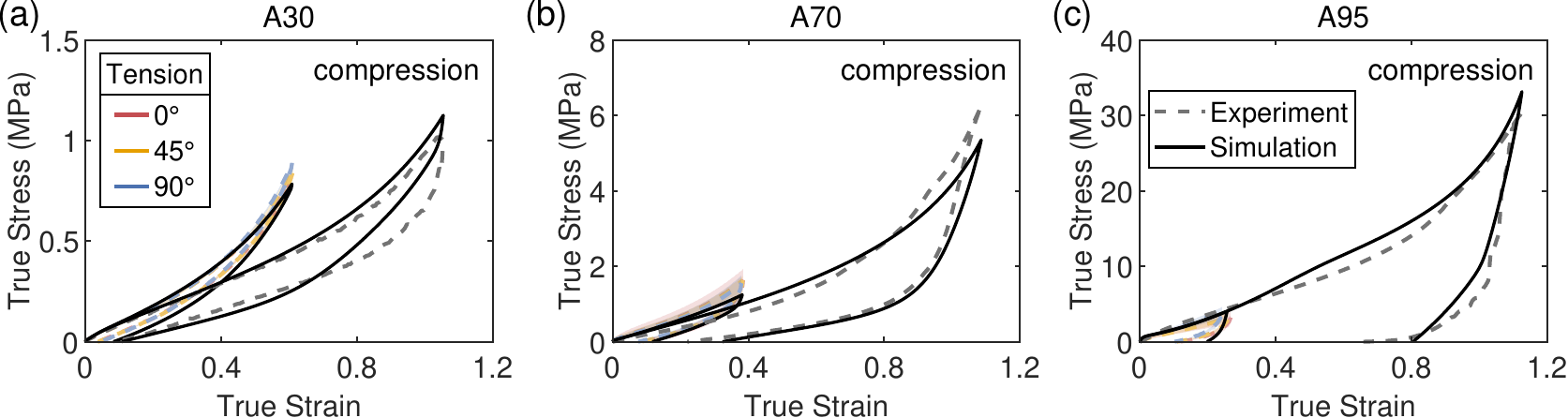}
    \caption{Predicted and measured uniaxial tension and compression of (a)~A30, (b)~A70 and (c)~A95 at $0.01~\mathrm{s^{-1}}$. Dashed curves, experiment; solid curves, model with the properties from uniaxial compression. The tension experiment is shown for specimens printed at $0^\circ$ (along $X$), $45^\circ$ and $90^\circ$ (along $Y$), each the mean of the tests at that orientation with their range shaded; the compression experiment is the calibration test.}
    \label{fig:tension_appendix}
\end{figure}

Uniaxial tension true stress-strain behavior is compared to uniaxial compression for A30, A70 and A95 at a strain rate of $0.01~\mathrm{s^{-1}}$. We note that the tensile strain achievable before failure in A70 and A95 is limited, and tests were then conducted to strains slightly below failure to show loading and unloading. The small strain response of uniaxial tension and compression overlap as expected and as predicted by the model. State of deformation dependence due to evolving molecular orientation is apparent at larger strain. A30 shows the clear state of deformation dependence of tension vs compression at large strains, including loading and unloading,  and is well captured by the model.  A70 begins to show this and is also captured by the model. A95 was not able to be tested to a large enough strain in tension to reveal any state of deformation dependence – the model did capture the behavior up to the given strain while overpredicting the immediate residual strain upon unloading.

\bibliographystyle{elsarticle-num-names}
\bibliography{references}

\end{document}